\documentclass[twocolumn]{aastex631}
\usepackage{amsmath}
\usepackage{float}
\usepackage{graphicx}
\usepackage{appendix}
\usepackage{xcolor}

\begin{document}

\title{Constraining the abundance of spinning deformed Galactic compact objects with continuous gravitational waves}


\author[0009-0001-2695-3622]{Gopalkrishna Prabhu}
\thanks{\text{Equal contribution from both authors.}}
\affiliation{Inter University Centre for Astronomy and Astrophysics, Post Bag 4, Ganeshkhind, Pune - 411007, India}


\author[0000-0003-0067-346X]{Aditya Kumar Sharma}
\thanks{\text{Equal contribution from both authors.}}
\affiliation{International Centre for Theoretical Sciences, Tata Institute of Fundamental Research, Bangalore 560089, India}

\author[0000-0002-6602-3913]{R. Prasad}
\affiliation{International Centre for Theoretical Sciences, Tata Institute of Fundamental Research, Bangalore 560089, India}

\author[0000-0001-5318-1253]{Shasvath J. Kapadia}
\affiliation{Inter University Centre for Astronomy and Astrophysics, Post Bag 4, Ganeshkhind, Pune - 411007, India}

\begin{abstract}
Galactic spinning compact objects (COs) with non-zero ellipticity are expected to be sources of continuous gravitational waves (CGWs). Certain classes of hypothetical COs, such as neutron stars with quark cores (hybrid stars), and quark stars, are thought to be capable of sustaining large ellipticities from theoretical considerations. Such exotic COs (eCOs) with large ellipticities and spins should produce CGWs detectable by the current LIGO-Virgo-Kagra GW detector network. Since no detections for CGWs, from searches in LIGO-Virgo data, have so far been reported, we place constraints on the abundance of highly elliptical, rapidly spinning, eCOs in our Galaxy. We formulate a Bayesian framework to place upper limits on the number count $N_{tot}$ of highly deformed Galactic eCOs. We divide our constraints into two classes: an ``agnostic'' set of upper limits on $N_{tot}$ evaluated on a CGW frequency and ellipticity grid that depend only on the choice of spatial distribution of COs; and a model-dependent set that additionally assumes prior information on the distribution of frequencies. We find that COs with ellipticities $\epsilon \gtrsim 10^{-5}$ have abundance upper limits at $90\%$ confidence, of $N_{tot}^{90\%} \lesssim 100$, and those with $\epsilon \gtrsim 10^{-6}$ have $N_{tot}^{90\%} \lesssim 10^4$. We additionally place upper-limits on the ellipticity of Galactic COs informed by our choices of spatial distributions, given different abundances $N_{tot}$.

\end{abstract}




\section{Introduction} \label{sec:intro}
The LIGO-Virgo-Kagra (LVK) network \citep{AdvLIGOdetector, AdvVIRGOdetector, KAGRAdetector} of gravitational-wave (GW) detectors has completed three observing runs (O1, O2, O3) \citep{LIGOScientific:2021djp}. These have produced $\sim 100$ detections, all of which correspond to compact binary coalescence (CBC) events. The majority of these are binary black hole mergers (BBHs), although binary neutron star (BNS) \cite{LIGOScientific:2017vwq, LIGOScientific:2020aai} and neutron-star black hole (NSBH) events have also been observed \citep{LIGOScientific:2021qlt}. 

The ongoing\footnote{at the time of writing} O4 could triple the number of detections reported so far, thus firmly establishing the arrival of GW astronomy. Indeed, even the existing list of detections has provided a wealth of scientific riches. These include unique tests of general relativity \citep{LIGOScientific:2021sio}, inference of population properties of BBHs that merge within Hubble time \citep{KAGRA:2021duu}, distance-ladder-independent measurements of the Hubble constant \citep{LIGOScientific:2021aug}, an understanding of the provenance of short-gamma-ray bursts and kilonovae \citep{LIGOScientific:2017ync}, and probes of ultra-dense matter via constraints on NS equations of state \citep{LIGOScientific:2018cki}. 

Another class of GWs, produced by spinning NSs (or possibly other hypothetical compact objects), are also expected to be detected with improved detector sensitivity, although no such detection has so far been reported \citep{KAGRA:2022osp, KAGRA:2022dwb, LIGOScientific:2021quq}. Spinning NSs with axial asymmetry as quantified by the ellipticity parameter $\epsilon$, will have a time-varying mass-quadrupole moment which in turn results in the production of GWs \citep[see, e.g., ][]{Bonazzola:1995rb}. However, unlike CBCs which are transient events with in-band durations of $O(\mathrm{min})$ at most in the current LVK network, GWs from NSs will be persistent events. Such continuous (C) GWs are expected to maintain a near-constant frequency, although their amplitudes, even for Galactic sources, are four orders of magnitude (or more) smaller than typical transient CBC sources detectable by the LVK network \citep[see, e.g.][]{Riles:2022wwz}. However, their enormous in-band duration could mitigate their small amplitudes by accumulating signal-to-noise ratio (SNR) across several cycles \citep[see, e.g.][]{jaranowski1998}.

The detectability of CGWs from Galactic COs crucially depends on the spin-frequencies of the COs, their ellipticities, and their distances from the Earth \citep[see, e.g][and references therein]{Piccinni:2022vsd}. Current upper limits on ellipticities of NSs \footnote{Or, for that matter, any CO with comparable moments of inertia} as a function of frequency, within the frequency band of the LVK network, from a directed search towards the center of the Galaxy, spans $\epsilon \in [10^{-7}, 1]$ \citep{KAGRA:2022osp}. These upper limits are expected to improve significantly with increasing observing time as well as improved sensitivity of future observing runs and future generation detectors such as Cosmic Explorer (CE) \citep{Reitze:2019iox} and Einstein Telescope (ET) \citep{punturo2010}.

Several mechanisms have been proposed that could impart finite ellipticities to NSs \citep[see, e.g.][]{glampedakis2018}, with an upper limit that could be as large as $\sim 7 \times 10^{-6}$ \citep{morales2022}. However, even such large ellipticities are likely to go undetected with current detector sensitivities if their spins are too small, or their distances too large. On the other hand, hypothetical exotic COs (eCOs) such as NSs with quark cores (hybrid stars) \citep{Owen:2005fn, haskell2007, glampedakis2012}, and quark stars \citep{Owen:2005fn}, are thought to be able to sustain ellipticities as large as few times $10^{-4}$, that could in principle enable the detection of such eCOs, provided they are sufficiently abundant in the Galaxy.

Conversely, the reported non-detection of CGWs from all-sky searches \citep{KAGRA:2022dwb, steltner2023deep, dergachev2023frequency, dergachev2024early} in LV's O1, O2, O3 data, has strong implications for the abundance of Galactic eCOs. In this {\it letter}, we formulate a Bayesian framework to constrain their number count. Specifically, from the CGW (frequency-dependent and sky-averaged) amplitude upper limits \citep{upperlimitdata}, and assuming that the eCOs are Poisson distributed, we construct a Bayesian posterior distribution on the number of COs ($N_{tot}$) from which we place upper limits on the number count at $90\%$ confidence. 

We present two classes of constraints. The first are ``agnostic'' constraints -- these are upper limits on $N_{tot}$ on a grid of ellipticities and CGW frequencies, assuming well-motivated spatial distributions of COs. Different portions in this parameter space pertain to allowed parameters for specific eCOs, viz., hybrid stars and quark stars. We also highlight the portion of the parameter space that pertains to highly elliptical NSs allowed under certain models.

The second are ``model-dependent'' constraints. These, assume prior information on the distribution of spins (motivated by those recorded in the ATNF catalog \citep{manchester2005}). Both classes of constraints strongly disfavor COs with ellipticities $\epsilon \gtrsim 10^{-5}$, with number count upper limits of $N_{tot} \lesssim 100$. Even ellipticities as low as $10^{-6}$, theoretically allowed for NSs, are scarce, with $N_{tot} \lesssim 10^4$. 

We additionally place spatial-distribution-informed upper limits on the ellipticity, assuming CO number counts of $10^{7,8,9}$ \citep{treves2000isolated}, consistent with the expected abundance of Galactic NSs. Our most conservative estimate yields $\epsilon \lesssim 1.4 \times  10^{-7}$.

The rest of the paper is organized as follows. Section~\ref{sec:method} provides the relevant equations pertaining to the CGW amplitude, motivates the possibility of the existence of high ellipticity eCOs, describes the astrophysically motivated spatial distribution of COs, and delineates the Bayesian framework used to constrain $N_{tot}$ and $\epsilon$. Section~\ref{sec:results}, presents both the ``agnostic'' and model-dependent results mentioned above. Section~\ref{sec:conclusions} summarizes the work and suggests scope for future work.

\section{Method}\label{sec:method}
\subsection{Continuous Gravitational-Waves from Spinning Compact Objects}

A spinning CO, such as an NS, with axial asymmetry, can be modeled as a triaxial ellipsoid. The asymmetry is typically quantified by the equatorial ellipticity parameter $\epsilon$ as follows:

\begin{equation}
    \epsilon = \frac{|I_{xx} - I_{yy}|}{I_{zz}}
\end{equation}

where $I_{zz}$ is the moment of inertia along the axis of rotation, and $I_{xx}, I_{yy}$ are moments of inertia in the plane perpendicular to the principal axis.

The amplitude of the corresponding GWs is given by:
\begin{equation}
    h_0 = \frac{16\pi^2 G}{c^4}\frac{\epsilon I_{zz}f^2_\star}{r}
    \label{amplitude}
\end{equation}
where $f_{\star}$ is the rotation frequency of the CO, and $r$ is the distance separating the CO from the Earth. The polarizations can then be written as:
\begin{equation}
    h_{+} = A_{+}h_0\sin(\omega t + \phi),~
    h_{\times} = A_{\times}h_{0}\cos(\omega t + \phi)
\end{equation}
with $\omega = 2\pi f$, $A_{+} = \sin\chi\frac{1 + \cos^2\iota}{2}$, $A_{\times} = \sin\chi\cos\iota$, and $\phi$ is a constant phase offset. Here, $\iota$ is the inclination angle between the axis of rotation of the CO and the observer's line of sight, and $\chi$ is the wobble angle between the axis of rotation and the principal axis. For the GW frequency, we consider the dominant second harmonic  $f_{GW} = 2f_{\star}$ throughout this work. 

The strain amplitude $h_0$ is orders of magnitude smaller than the transient events detected so far but stays in the frequency band of the detector for a very long period. The signal is integrated over a longer duration to build SNR for such sources.  There have been searches for such continuous GW signals through the three observing runs of LIGO. The all-sky search performed in O3 by LVK \citep{allskyo3} uses FrequencyHough \citep{frequencyhough}, SkyHough \citep{skyhough}, Time-Domain F-Statistic \citep{timedomainfstat1,timedomainfstat2}, and SOAP \citep{soap} to search for CGWs from isolated NSs in the $10 \textendash 2048$ Hz frequency range and frequency derivative range of $-10^{-8} \textendash  10^{-9}$. \footnote{The parameter range for individual searches varies across search pipelines. Please refer to \citep{allskyo3} for further details on the search parameter space.} In addition to these searches by LVK collaboration, other groups, e.g. Einstein@Home \citep{steltner2023deep}, Falcon search \citep{dergachev2023frequency} have also independently searched for CGWs in O3 data. Given the lack of any detections, by searches within and outside the LVK collaboration, upper limits it is natural to place upper limits on the strain $h_0$.  We use the $95 \%$ confidence level upper limits from the latest LVK O3 FrequencyHough all-sky search to constrain the population of eCOs in this work.

\begin{figure}[H]
    \centering
    \includegraphics[scale=0.24]{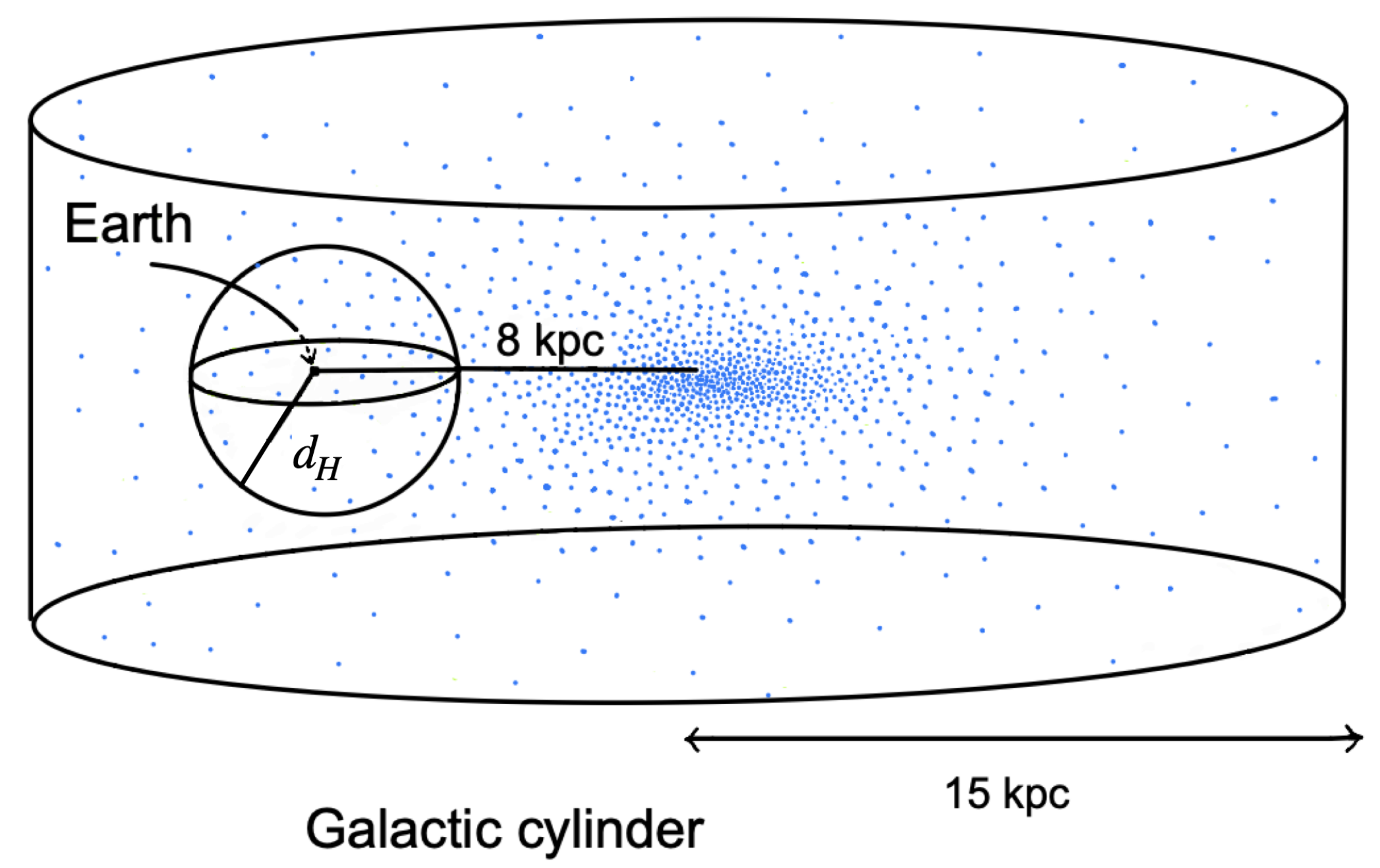}
    \caption{Illustration of the mean number of Galactic COs that lie within the horizon sphere with radius $r = d_H(\epsilon, f_{GW})$. This number is proportional to the probability volume of the 3D spatial distribution that lies within this sphere.}
    \label{fig:illustration}
\end{figure}

\begin{figure*}[h!tbp] 
\centering
 \includegraphics[scale=0.6]{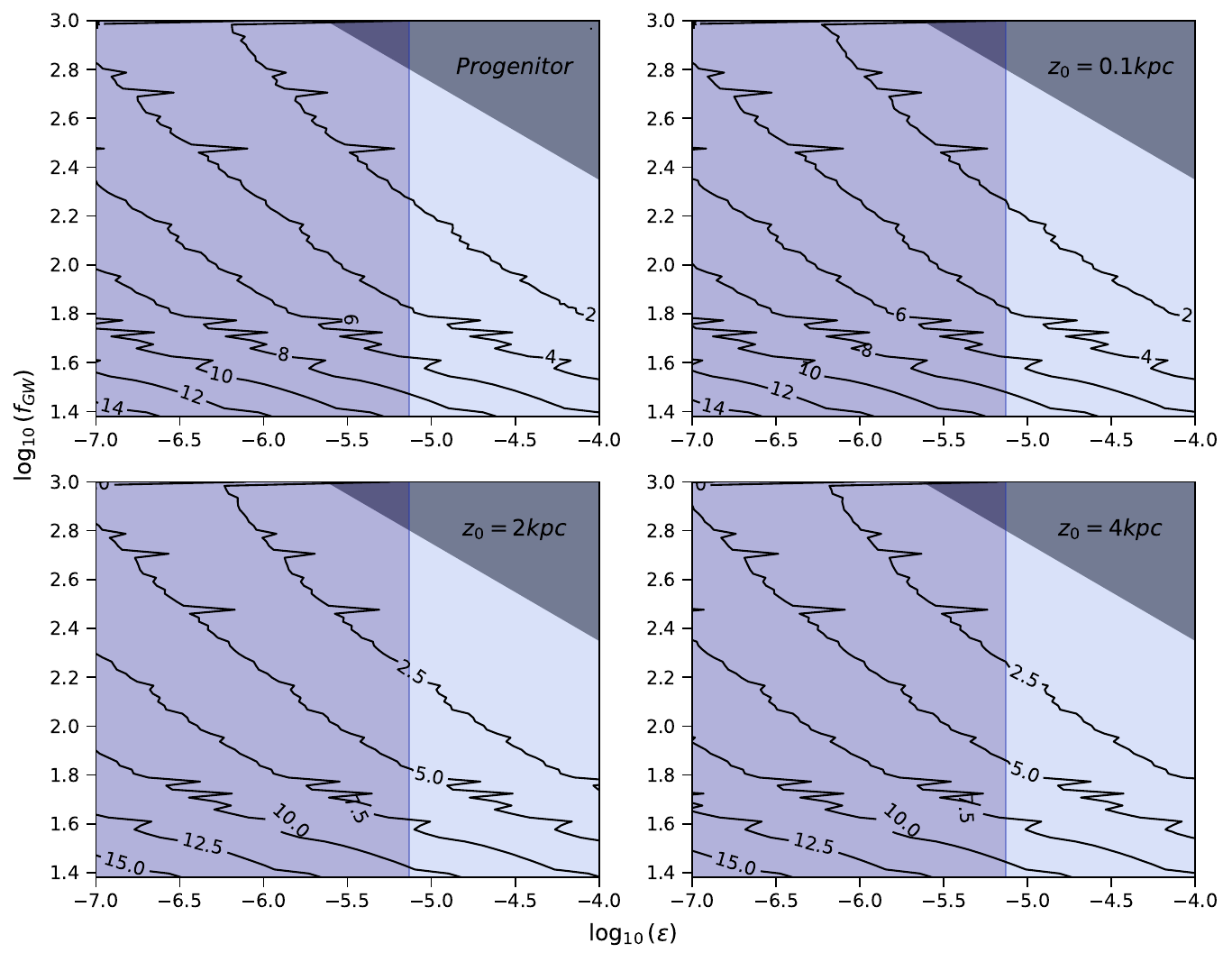} 
\caption{Two-dimensional density plots for $\log_{10}(N_{tot}^{90\%})$ as a function of frequency $f_{GW}$ and ellipticity $\epsilon$. Each panel corresponds to one choice of spatial distribution, viz., the Progenitor \citep{Paczynski} and the $3$ ``Reed'' \citep{Reed} distributions. The solid black lines are contours of constant $\log_{10}N_{tot}^{90\%}$. The dark grey shade on the top right corners of the panels are regions of the parameter space not probed by the FrequencyHough method. The vertical line demarcates the maximum ellipticity the NS crust can support \citep{morales2022}. These plots give the most model-agnostic results among those presented in this work since the only model assumption comes from the choice of spatial distributions. It is interesting to note that with varying spatial distributions, the number count, for a given frequency and ellipticity, does not change by more than about an order of magnitude.} 
\label{fig:2dplot} 
\end{figure*}

\subsection{Maximum ellipticities of compact objects} 
In addition to conventional NSs, it is theoretically plausible for hybrid stars, containing deconfined quark matter within their central regions, and quark stars, which are entirely made of quark matter, to exist \citep{weber2007pulsars,glendenning2012compact}. CGWs emission depends on the emitting star's ellipticity, which is closely linked to the type of compact object and the mechanism causing its structural deformation. If a neutron star's magnetic field isn't aligned with its axis of spin, it could induce shape deformations, which could lead to finite ellipticity \citep{chandrasekhar1953problems,ferraro1954equilibrium,gal1984gravitational}. This magnetic ellipticity changes as a square of the magnetic field \citep{bonazzola1996gravitational,konno2000flattening} and also depends on how strong the poloidal and toroidal components are in mixed-field configurations \citep{
haskell2008modelling, mastrano2011gravitational}. According to standard models \citep{mastrano2011gravitational,lasky2015gravitational}, 

\begin{equation}\label{epsilon-bfield-ns}
\epsilon \approx 4.5 \times 10^{-7} \left( \frac{B_{p}}{10^{14}G} \right)^{2} \left(1- \frac{0.389}{\Lambda} \right),
\end{equation}

where $B_p$ is the poloidal component of the surface field, and $ \Lambda$ is the ratio of poloidal energy to total magnetic energy. Even the powerful magnetic fields ($10^{14}-10^{15} G$) reported in some NSs \citep{olausen2014mcgill} cannot produce substantial ellipticities through structural deformations. However, in hybrid stars, a superconducting quark phase can lead to significant ellipticity through the effects of color-magnetic stresses in the core \citep{glampedakis2012}. This ellipticity exhibits a linear relationship with the magnetic field and is given by \citep{lasky2015gravitational},
\begin{equation}\label{epsilon-bfield-qs}
\epsilon \approx 1.2 \times 10^{-5} \frac{<B>}{10^{14} \text{ G}}.
\end{equation}

In NSs that have undergone a history of accretion, the accreted matter can get funneled by the magnetic field, which can cause mountains to form \citep{brown1998ocean,melatos2005gravitational}. Also, the uneven temperatures in the star's interior can induce deformations \citep{haensel1990non,ushomirsky2000deformations}. These deformations are sustained by the crustal strains and can be quantified by the maximum stress the crust can withstand before breaking. This ellipticity is considerably larger for quark stars composed of solid strange matter \citep{xu2003solid}, with values as large as $\epsilon = 10^{-4}$, and the hybrid stars up to $10^{-5}$ \citep{Owen:2005fn,haskell2006mountains, haskell2007, glampedakis2018}. Maximum elastic quadrupolar deformations calculated in \cite{nathan2013} suggest that certain hybrid stars and quark stars could sustain ellipticities up to $10^{-3}$ and $10^{-1}$ respectively for a fiducial moment of inertia of $10^{38} \mathrm{kg \, m^2}$. 

On the other hand, typical NSs are thought to support a maximum ellipticity of about a few times $10^{-6}$ \citep{morales2022}. Regardless of the ellipticity-producing mechanism described in different theoretical models (magnetic fields, crustal strains, or color-magnetic stresses), very large ellipticities are theoretically permissible for COs (and eCOs in particular), allowing us to examine their population within our Galaxy. 

\subsection{Spatial distributions of Galactic neutron stars} \label{spatial distribution}

To establish a connection between the upper limits on strain amplitude and the abundance of COs, we need assumptions on the spatial distribution of these objects. Thus, we explore various spatial distribution models for COs, with the simplest being the ``Progenitor" distribution proposed by \citep{Paczynski}. This model assumes that COs follow the same spatial distribution as their progenitor stars. For these Progenitor models, the probability distribution in the cylindrical coordinates centered at the Galactic center is given by:
\begin{align}
    \frac{dP}{dR} \, &= \, a_R \, \frac{R}{R_0^2} \,  \exp \left(-\frac{R}{R_0}\right)  \\
    \frac{dP}{dz} \, &= \, \frac{1}{2 z_0} \, \exp \left(- \frac{|z|}{z_0}\right)    
\end{align}
$a_R$ is the normalization constant for the radial distribution. The scales $R_0$ and $z_0$ for this progenitor model are set to be $4.5$ kpc and $0.07$ kpc, respectively.

The assumption of COs inheriting the spatial distribution of their progenitor stars is valid only when the natal kicks these objects receive are minimal or when they are relatively young remnants of stellar evolution. This simplifying assumption should be noted when interpreting results pertaining to the Progrenitor distribution. 

Studies like \cite{Reed} have proposed spatial distributions with different scale heights $z_0$ so as to take the birth kicks into account, while also evolving the initial (progenitor) spatial distribution within the Galactic potential across billions of years. We consider their proposed distribution in our study, with scale factors set to be, $z_0 \, = \, 0.1 \text{kpc}, \, 2 \, \text{kpc, and } 4 \, \text{kpc}$.  Additionally, for these ``Reed" \citep{Reed} models, the radial distribution is taken to be a Gaussian, with the standard deviation, $\sigma_R$, chosen to be $4.5$ kpc.

\begin{align}
    \frac{dP}{dR} \, &= \, \frac{R}{\sigma_R^2} \, \exp \left(- \frac{R^2}{2 \sigma_R^2}\right)
\end{align}

All of these models assume azimuthal symmetry. From these, we construct the 3-D spherical distribution centered on Earth, by using appropriate coordinate transformation and the Jacobian associated with this transformation. In the next subsection \ref{bayesian}, where we estimate the abundance of Galactic COs based on the non-detection of CGWs using a Bayesian framework, these 3-D spatial distributions are used to find the average fraction of COs that lie within the detection horizon (the ``horizon sphere'') of the O3 all-sky searches.

\subsection{Bayesian framework} \label{bayesian}

\begin{figure*}[h!tbp] 
\centering
\includegraphics[scale=0.4]{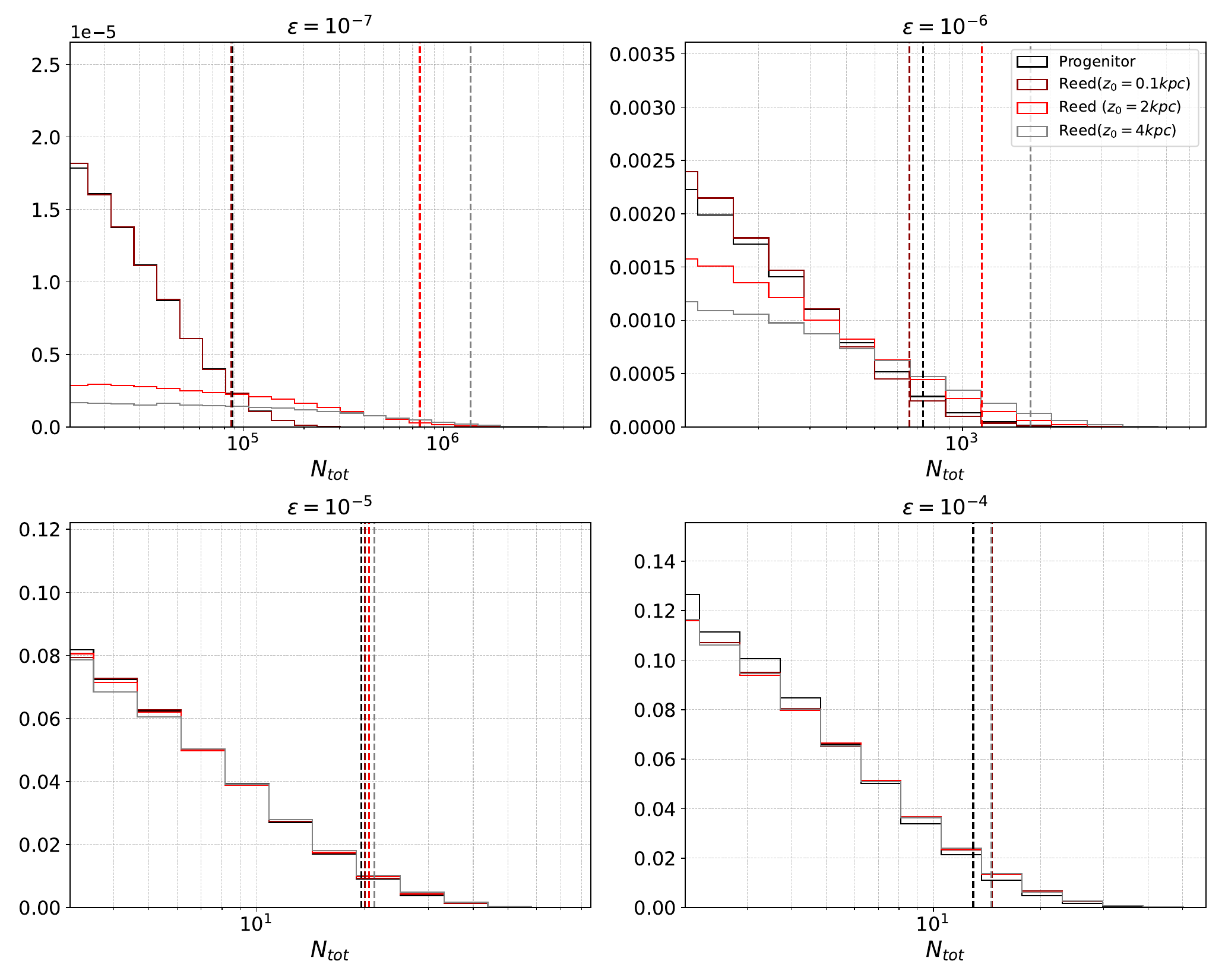} 
\caption{Histograms of samples drawn from the posterior on $N_{tot}$ (cf. Eq.~\ref{Ntotalprobability}), marginalized with respect to the ATNF-like frequency distribution. The progenitor and $3$ ``Reed'' distributions are considered, and the histograms are evaluated for four choices of ellipticities $\epsilon = 10^{-7,-6,-5,-4}$. The upper limits for $\epsilon = 10^{-5,-4}$ are less than $100$, suggesting that the abundance of such high-ellipticity objects in the Galaxy is close to being ruled out. On the other hand, the abundance of $\epsilon = 10^{-6}$ is constrained to less than $10^4$, which is four to five orders of magnitude lower than the abundance of Galactic NSs. The weakest constraints in this plot pertain to $\epsilon = 10^{-7}$, with a number count of less than $10^7$, which is still one to two orders of magnitude smaller than the Galactic NS abundance.} 
\label{fig:NtotMargDist} 
\end{figure*}

The strain upper limits ($h_0^{\mathrm{upper}}$) placed by the CGW searches can be converted to upper limits on the detectable range, $d_{H}$, given an ellipticity ($\epsilon$), frequency $f_{\mathrm{GW}}$ and moment of inertia $I_{zz}$. Using Eq \eqref{amplitude}, we get:
\begin{equation}
    d_{H}=\frac{4\pi^2 G I}{c^4}\frac{f_{GW}^2\epsilon}{h_0^{\mathrm{upper}}(f_{GW})}
    \label{horizon}
\end{equation}
where $f_{GW} = 2f_{\star}$ has been assumed. We pin the moment of inertia to a fiducial value of $10^{38} \, \mathrm{kg \, m^2}$.
 
 Assuming a spatial distribution of COs and given the horizon distance ($d_H$), we can find the mean fraction of the total number $N_{tot}$ of COs that lie within a sphere centered at the Earth with radius $r = d_{H}$. Given $N_{tot}$, this fraction, which we call $\alpha (\epsilon, f_{GW})$ gives us the fraction of detectable COs, and is found by integrating the 3-D spatial distribution over the horizon sphere (see Fig.~\ref{fig:illustration}).: 
 \begin{equation}
     \alpha (\epsilon,f_{gw})=\int_0^{d_H}\int _0^{\pi}\int_0^{2\pi}p(r,\theta,\phi) \, dr \, d\theta \, d\phi
     \label{mean_fraction}
 \end{equation}
Here, $r, \theta, \phi$ are the usual spherical polar coordinates centered on the Earth, and $p(r, \theta, \phi)$ is the spatial distribution described in the previous subsection in these coordinates.

The actual number $n$ that lies within the horizon sphere, with ellipticity $\epsilon$ and frequency $f_{\star} = f_{GW}/2$, follows a Poisson distribution with a mean $\alpha N_{tot}$: \footnote{Strictly, the detectable COs should conform to a binomial distribution. In the large $N_{tot}$ limit, this binomial distribution can be well approximated to be a Poisson distribution with a mean of successful trials, i.e. total detectable events, $\alpha N_{tot}$.}
%
\begin{equation} 
       p\left(n|N_{tot},\alpha (\epsilon,f_{GW})\right)= \frac{(\alpha N_{tot})^n \exp(-\alpha N_{tot})}{n!}
       \label{posterior_ndet}
\end{equation} 
Asserting that exactly $n = 0$ COs were detected, a simple application of Bayes' rule yields the following:
\begin{equation} 
      p(N_{tot}|n=0,\alpha)  \propto \exp(-\alpha N_{tot})p(N_{tot})
      \label{Ntotalprobability}
\end{equation}
where $p(N_{tot})$ is a prior, which we consider uniform.

This prescription allows us to construct the posteriors on $N_{tot}$ and thus extract upper limits on the abundance of COs with specific ellipticities and frequencies. Assuming that the frequency distribution of COs follows the ATNF catalog, a marginalized posterior on $N_{tot}$ whose shape depends only on ellipticity, can also be constructed. 

%

One can equivalently evaluate a spatial-distribution-informed posterior on the upper limit for ellipticity. Assuming that ellipticities are log-uniformly distributed between $[\epsilon_{min}, \epsilon_{max}]$, and that the frequencies follow the ATNF catalog, we can construct a posterior distribution on $\epsilon_{max}$ using the following procedure. Similar to Eq(\ref{mean_fraction}), the mean fraction of detectable COs given a fixed $N_{tot}$ and a log-uniform distribution on $\epsilon \in [\epsilon_{min},\epsilon_{max}]$ is given by: 

\begin{align}
    \alpha(\epsilon_{max}) &= \int df p(f)\int d\epsilon p(\epsilon|\epsilon_{max}) \nonumber \\ & \int_0^{d_H(\epsilon,f_{GW})}\int_0^{\pi}\int_0^{2\pi}d\phi d\theta dr \, p(r,\theta,\phi)
    \label{AlphaEpsMax}
\end{align}

We adopt a Monte-Carlo approach where samples are first drawn from the chosen spatial distribution, the ATNF catalog frequency distribution, and the ellipticities up to an $\epsilon_{max}$. The fraction of samples that lie within the horizon sphere is the desired $\alpha$. Repeating this exercise for a range of $\epsilon_{max}$ enables us to construct $\alpha(\epsilon_{max})$. \footnote{The Monte-Carlo approach has its own associated Poisson errors, which have not been accounted for. These errors increase with decreasing ellipticity, where relatively smaller fractions of samples lie within the horizon sphere.}

Thus, analogous to Eq(\ref{posterior_ndet}), the likelihood function of $\alpha(\epsilon_{max})$ can now be written as:
\begin{equation}
    p\left(n|N_{tot},\alpha(\epsilon_{max})\right)=\frac{\left(\alpha(\epsilon_{max})N_{tot}\right)^n exp\left(-\alpha(\epsilon_{max})N_{tot}\right)}{n!}
    \label{LikelihoodAlphaEpsMax}
\end{equation}
Re-parameterization of this likelihood function ($\alpha(\epsilon_{max})\rightarrow \epsilon_{max}$) allows us to write the likelihood as \footnote{Note that changing variables in the likelihood function does not require a Jacobian. However, if we had chosen to set a prior on $\alpha(\epsilon_{max})$, we would need to transform that prior into one for $\epsilon_{max}$ using an appropriate Jacobian for the variable transformation. Instead, we opt for a uniform prior directly on $\epsilon_{max}$, rather than setting a prior on $\alpha(\epsilon_{max})$.}:
\begin{equation}
p(n|N_{tot},\alpha(\epsilon_{max}))=p(n|N_{tot},\epsilon_{max})
\end{equation}
Now, using Bayes' theorem, and asserting as before that exactly $n=0$ detections were made, we acquire the posterior on $\epsilon_{max}$ (cf. Eq.~\ref{LikelihoodAlphaEpsMax}):
%
\begin{align}
    p(\epsilon_{max}|n=0,N_{tot})&\propto \nonumber p(n=0|N_{tot},\epsilon_{max})p(\epsilon_{max}) \\& 
    \propto \exp(-\alpha(\epsilon_{max}) N_{tot})
    \label{epsilonmax}
\end{align}

%


\section{Results}\label{sec:results}

We estimate upper limits, at $90\%$ confidence, on the abundance of COs, as a function of ellipticity $\epsilon$ and GW frequency $f_{GW}$. To do so, we first calculate the detectability range $d_H(\epsilon, f_{GW})$ over a frequency-ellipticity grid (cf. Eq.~\ref{horizon}). We use Eq.~\ref{mean_fraction} to evaluate the mean detectable fraction $\alpha(\epsilon, f_{GW})$ over the grid, for each of the spatial distributions (the Progenitor and the three ``Reed'' distributions with different scale heights). The posteriors on the abundance $N_{tot}$, $p(N_{tot}|n=0, \alpha(\epsilon, f_{GW}))$, are then acquired from Eq.~\ref{Ntotalprobability}.

We sample $p(N_{tot}|n=0, \alpha(\epsilon, f_{GW}))$ posteriors and calculate the $90\%$ quantile, $N_{tot}^{90\%}$, across the frequency-ellipticity grid. We plot these upper limits in Figure~\ref{fig:2dplot}. Each panel in the figure corresponds to one choice of spatial distribution. The contour lines of constant $\log_{10}(N_{tot}^{90\%})$ conservatively suggest a paucity of COs with large ellipticities, potentially (though not strictly) ruling them out. COs with ellipticities $> 10^{-4}$ and frequencies $> 100$ Hz have $N_{tot}^{90\%} \lesssim 100$, regardless of choice of spatial distribution. Largest allowed NS ellipticities \citep{morales2022} and frequency $\gtrsim 100$ Hz have $N_{tot}^{90\%} \lesssim 10^5$. Indeed, distributions with small scale heights give even more stringent constraints of $N_{tot}^{90\%} \lesssim 10^4$, thus strongly limiting the abundance of high-ellipticity Galactic millisecond pulsars.

Assuming that the frequency distribution $p(f_{\star})$ follows the ATNF catalog, we construct $p(N_{tot}|n=0, \alpha(\epsilon))$ by marginalising $\alpha(\epsilon, f_{GW}))$ with respect to this distribution. We plot the corresponding $N_{tot}$ posteriors for fixed ellipticities, and the four spatial distributions, in Figure~\ref{fig:NtotMargDist}. We demarcate the $N_{tot}^{90\%}$ upper limits for the chosen ellipticities and spatial distributions, for reference.

\begin{figure}
    \vspace{0.2 cm}
    \centering
    \includegraphics[scale=0.43]{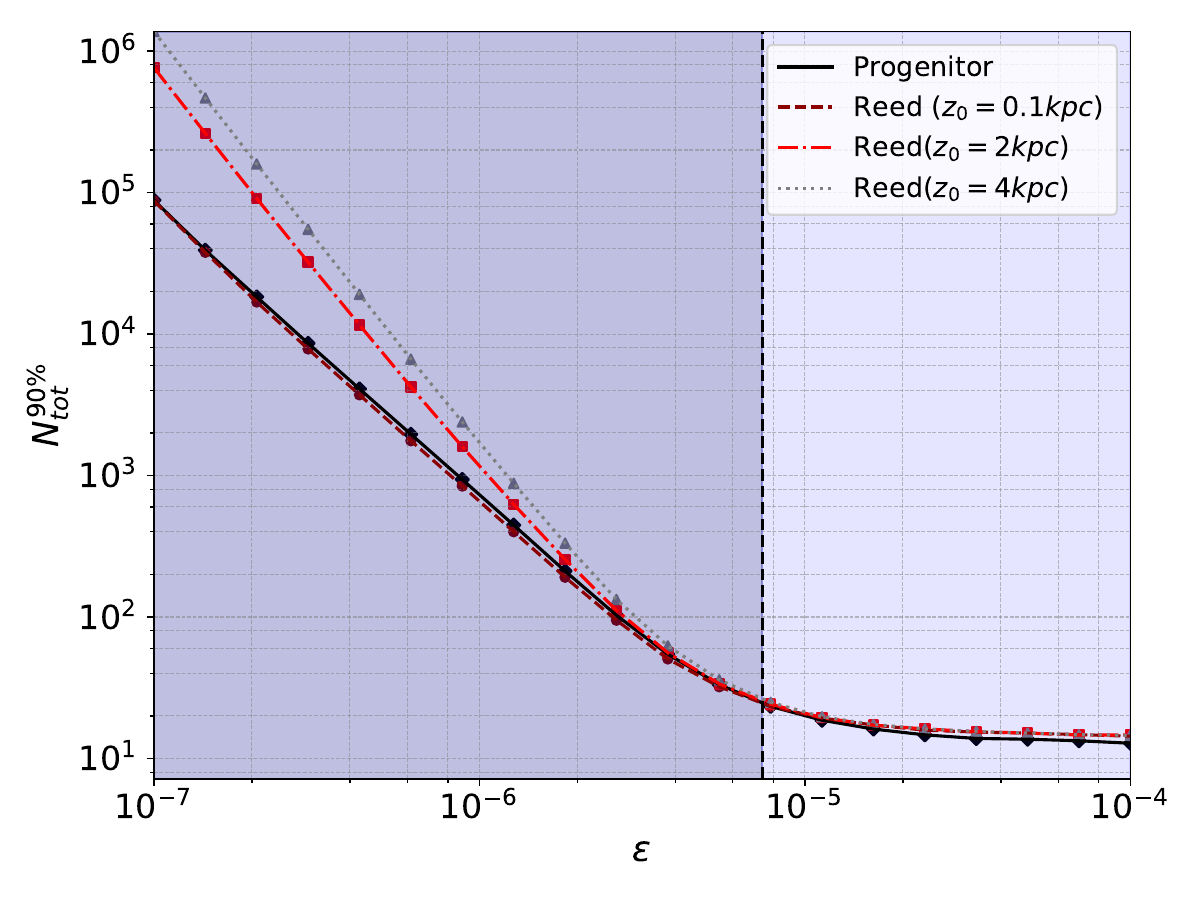}
    \caption{Upper limits $N_{tot}^{90\%}$, as a function of ellipticity $\epsilon$, acquired from the posterior distribution on $N_{tot}$ marginalized with respect to an ATNF-like frequency distribution. The vertical line is the same as in Figure~\ref{fig:2dplot}. $\epsilon > 10^{-5}$ yields $N_{tot}^{90\%} < 100$, suggesting that such ellipticities are virtually ruled out in the Galaxy, under the assumed spatial and frequency distributions. Moreover, there is also, a scarcity of high-ellipticity NSs that are theoretically allowed, with $N_{tot}^{90\%} < 10^4$ for $\epsilon > 10^{-6}$.}
    \label{fig:NtotFrequency}
\end{figure}

We additionally plot $N_{tot}^{90\%}$ as a function of $\epsilon$, for a range of $\epsilon$, in Figure~\ref{fig:NtotFrequency}. Once again, we find that ellipticities larger than $10^{-5}$ have a number count less than $100$, at $90\%$ confidence, and are therefore strongly disfavored. While these results present stringent constraints on the number count, they must be interpreted with some caution, because the ATNF catalog may not adequately represent the frequency distribution of Galactic COs. 

To estimate the upper limit on the ellipticity, $\epsilon_{max}$, of the Galactic COs, we assume a log-uniform distribution $\epsilon \in [\epsilon_{min}, \epsilon_{max}]$. We fix $\epsilon_{min}$ to $10^{-12}$, and construct a posterior distribution on $\epsilon_{max}$ as described in the previous section, using Eq.~\ref{epsilonmax}. We pick $N_{tot} = 10^{7,8,9}$, consistent with the general expectation of the abundance of neutron stars in the Galaxy \citep{treves2000isolated}. 

\begin{figure*}[t!]
    \centering
    \includegraphics[width=\textwidth]{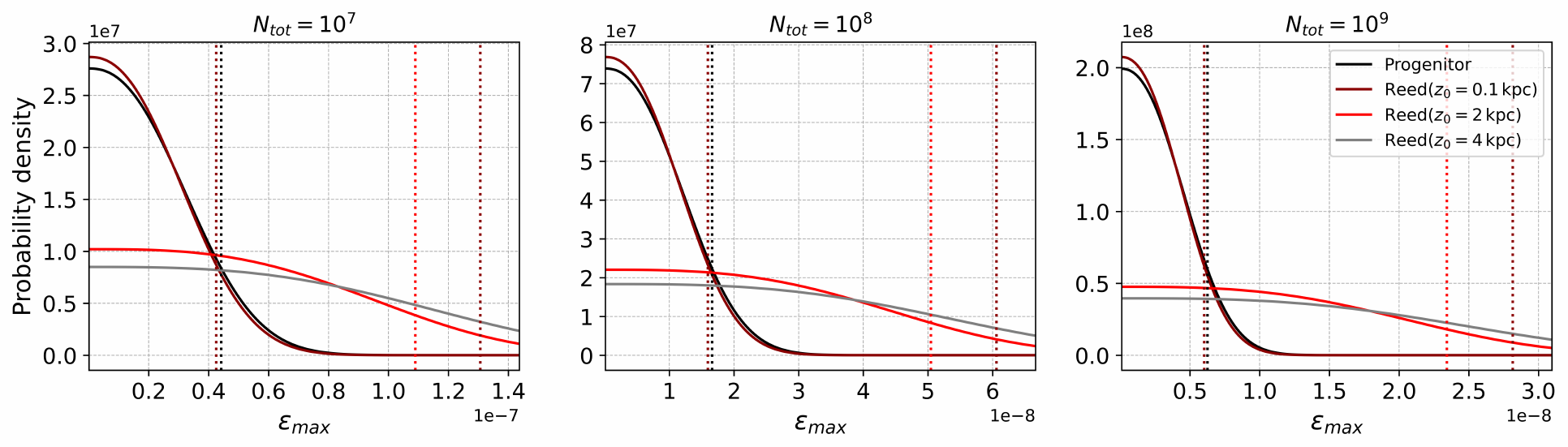}
    \caption{Posterior on $\epsilon_{max}$, assuming an ATNF-like frequency distribution, four spatial distributions, $N_{tot} = 10^{7,8,9}$, and a log-uniform ansatz on the ellipticity distribution with $\epsilon \in [\epsilon = 10^{-12}, \epsilon_{max}]$. Among all the cases considered, the most conservative upper limit is $\epsilon_{max} \sim 1.4 \times 10^{-7}$.}
    \label{fig:epsilonmax}
\end{figure*}

We plot $p(\epsilon_{max}|n = 0, N_{tot} = 10^{7,8,9})$ in Fig.~\ref{fig:epsilonmax} for the four choices of spatial distributions. We choose an uniformative prior on $\epsilon_{max}$ which we set to be uniform between $10^{-10}$ and $10^{-3}$. Upper limits on $\epsilon_{max}$, at $90\%$ confidence, are also demarcated. Our most conservative estimates suggest that Galactic COs have ellipticities $\lesssim 1.4 \times 10^{-7}$, strongly disfavoring eCOs with large ellipticities. This upper limit constraint appears to be slightly stronger than the frequency-dependent upper limits from searches directed toward the Galactic center. However, this improved upper limit is contingent on the choice of frequency and, to a lesser extent, the spatial distribution of COs, and should accordingly be interpreted. 

\section{Summary and Outlook}\label{sec:conclusions}

Spinning compact objects (COs) with a non-zero ellipticity ($\epsilon$) are expected to be sources of persistent GWs. The amplitude of these GWs depends linearly on the ellipticity and quadratically on the spin frequency. Various searches for these (C)GWs have been conducted in data from the first three observing runs of the LVK detector network. These searches produced a null result.  

We present a Bayesian framework to infer the abundance of Galactic COs from the non-detection of CGWs, and the frequency-dependent GW amplitude upper limits provided by the searches. We refer the reader to other formalisms, \citep{tenorio2023blindsearch} that discusses constraints on sub-kiloparsec CGW sources (e.g. Gravitars), \citep{wade} which provides prospective bound on ellipticity and magnetic field parameters with or without detection of CGWs, \citep{knispel2008} which discusses the expected CGW amplitudes from the galactic population of Gravitars and \citep{pagliaro2023continuous} which discusses the prospects of detecting standard NS's in current and future observing scenarios using realistic simulations.

We consider four astrophysically motivated spatial distributions of COs. The first follows the Galactic distribution of stars (``Progenitor'' distribution), while the other three are motivated from large-scale simulations that evolve an initial spatial distribution in the Galactic potential, across billions of years, and also incorporate natal kicks \citep{Paczynski, Reed}.

The mean number-fraction of COs that are detectable is proportional to the spatial distribution probability volume within the detection horizon sphere of radius $r = d_H(\epsilon, f_{GW})$. Assuming that the present distribution of COs across the Galaxy is one realization of the chosen spatial distribution, and follows Poisson statistics, we construct a posterior on the total number $N_{tot}$ of Galactic COs as a function of ellipticity and frequency. We evaluate the $90\%$ upper limits on $N_{tot}$, across a frequency-ellipticity grid. We also calculate these upper limits as a function of ellipticity alone, by marginalizing $\alpha(\epsilon, f_{GW}))$ with respect to an ATNF-like distribution of frequencies. We find that high-frequency ($> 100$ Hz) and high-ellipticity ($> 10^{-5} $) COs have $N_{tot}^{90\%} < 10^5 $, suggesting a paucity of such COs. 

Moreover, assuming a log-uniform distribution in ellipticities up to an $\epsilon_{max}$, and an $N_{tot} = 10^{7,8,9}$, in addition to the previous assumptions on spatial and frequency distributions, we determine a posterior on $\epsilon_{max}$. The most conservative $90\%$ upper limits suggest $\epsilon_{max} \lesssim 1.4 \times 10^{-7}$. This is a stronger (though model-dependent) upper limit than those inferred from searches targeted at the Galactic center \citep{KAGRA:2022osp}.

Our work implies a severe scarcity of high-ellipticity Galactic COs. Galactic solid strange quark stars with ellipticities of $\epsilon \sim 10^{-4}$ or higher, appear to be ruled out under our model assumptions. Hybrid stars with $\epsilon \gtrsim 10^{-5}$ have $N_{tot}^{90\%} \lesssim 10^2$, which is six to seven orders of magnitude smaller than the abundance of Galactic NSs. Even among Galactic NSs, those with $\epsilon \gtrsim 10^{-6}$, while permissible theoretically, are likely to be rare, with $N_{tot}^{90\%} \lesssim 10^4$. \footnote{Note that our number constraints are not strictly on the abundance of eCOs in general, but on those with significant deformations.}

It is interesting to probe the physical parameters of the NS with these constraints. A simple, back-of-the-envelope calculation using Eq.~\ref{epsilon-bfield-ns} suggests a scarcity of high-magnetic field NSs, with number counts for $B > 10^{14} G$ being $N_{tot}^{90\%} \lesssim 10^6$. The constraints are even more stringent for high-magnetic field hybrid stars, with $B > 10^{14} G$ yielding $N_{tot}^{90\%} \lesssim 10^2$ (cf. Eq.~\ref{epsilon-bfield-qs}). However, such statements need to be made more rigorous, since multiple physical mechanisms can impart an ellipticity, and not just magnetic fields. Moreover, spin-down effects from these mechanisms will also need to be accounted for. Using results from other searches for CGWs in O3 data \citep{steltner2023deep, dergachev2023frequency, dergachev2024early}, in tandem with the LVK's results, we plan to put constraints on said physical parameters of Galactic COs in future work.

\section{Acknowledgments}
We thank Rodrigo Tenorio and Nathan Johnson-McDaniel for their review and comments on the manuscript. We also thank Debarati Chatterjee, Suprovo Ghosh, Surhud More, Aditya Vijaykumar, Karl Wette, and members of the ICTS Astrophysical Relativity Group, for useful discussions. All computations were performed on the Sarathi Computing Cluster at IUCAA and the Alice Computing Cluster at ICTS-TIFR. AKS and RP  acknowledge the support of the Department of Atomic Energy, Government of India, under project No. RTI4001. 

This research has made use of data or software obtained from the Gravitational Wave Open Science Center (gwosc.org), a service of the LIGO Scientific Collaboration, the Virgo Collaboration, and KAGRA. This material is based upon work supported by NSF's LIGO Laboratory which is a major facility fully funded by the National Science Foundation, as well as the Science and Technology Facilities Council (STFC) of the United Kingdom, the Max-Planck-Society (MPS), and the State of Niedersachsen/Germany for support of the construction of Advanced LIGO and construction and operation of the GEO600 detector. Additional support for Advanced LIGO was provided by the Australian Research Council. Virgo is funded, through the European Gravitational Observatory (EGO), by the French Centre National de Recherche Scientifique (CNRS), the Italian Istituto Nazionale di Fisica Nucleare (INFN) and the Dutch Nikhef, with contributions by institutions from Belgium, Germany, Greece, Hungary, Ireland, Japan, Monaco, Poland, Portugal, Spain. KAGRA is supported by Ministry of Education, Culture, Sports, Science and Technology (MEXT), Japan Society for the Promotion of Science (JSPS) in Japan; National Research Foundation (NRF) and Ministry of Science and ICT (MSIT) in Korea; Academia Sinica (AS) and National Science and Technology Council (NSTC) in Taiwan.
\vspace{5mm}

%

\vspace{5mm}


\textit{Software}: \texttt{NumPy} \citep{vanderWalt:2011bqk}, \texttt{SciPy} \citep{Virtanen:2019joe}, \texttt{astropy} \citep{2013A&A...558A..33A, 2018AJ....156..123A}, \texttt{Matplotlib} \citep{Hunter:2007}, \texttt{jupyter} \citep{jupyter}, \texttt{pandas} \citep{mckinney-proc-scipy-2010}.





   


\bibliography{references}{}

\begin{thebibliography}{}
\expandafter\ifx\csname natexlab\endcsname\relax\def\natexlab#1{#1}\fi
\providecommand{\url}[1]{\href{#1}{#1}}
\providecommand{\dodoi}[1]{doi:~\href{http://doi.org/#1}{\nolinkurl{#1}}}
\providecommand{\doeprint}[1]{\href{http://ascl.net/#1}{\nolinkurl{http://ascl.net/#1}}}
\providecommand{\doarXiv}[1]{\href{https://arxiv.org/abs/#1}{\nolinkurl{https://arxiv.org/abs/#1}}}

\bibitem[{Aasi {et~al.}(2014)Aasi, Abbott, Abbott, Abbott, Abernathy, Accadia, Acernese, Ackley, Adams, Adams, Addesso, Adhikari, Affeldt, Agathos, Aggarwal, Aguiar, Ain, Ajith, Alemic, Allen, Allocca, Amariutei, Andersen, Anderson, Anderson, Anderson, Arai, Araya, Arceneaux, Areeda, Aston, Astone, Aufmuth, Aulbert, Austin, Aylott, Babak, Baker, Ballardin, Ballmer, Barayoga, Barbet, Barish, Barker, Barone, Barr, Barsotti, Barsuglia, Barton, Bartos, Bassiri, Basti, Batch, Bauchrowitz, Bauer, Behnke, Bejger, Beker, Belczynski, Bell, Bell, Bergmann, Bersanetti, Bertolini, Betzwieser, Beyersdorf, Bilenko, Billingsley, Birch, Biscans, Bitossi, Bizouard, Black, Blackburn, Blackburn, Blair, Bloemen, Blom, Bock, Bodiya, Boer, Bogaert, Bogan, Bond, Bondu, Bonelli, Bonnand, Bork, Born, Borkowski, Boschi, Bose, Bosi, Bradaschia, Brady, Braginsky, Branchesi, Brau, Briant, Bridges, Brillet, Brinkmann, Brisson, Brooks, Brown, Brown, Brückner, Buchman, Bulik, Bulten, Buonanno, Burman, Buskulic, Buy, Cadonati, Cagnoli,
  Bustillo, Calloni, Camp, Campsie, Cannon, Canuel, Cao, Capano, Carbognani, Carbone, Caride, Castiglia, Caudill, Cavalier, Cavalieri, Celerier, Cella, Cepeda, Cesarini, Chakraborty, Chalermsongsak, Chamberlin, Chao, Charlton, Mottin, Chen, Chen, Chincarini, Chiummo, Cho, Chow, Christensen, Chu, Chua, Chung, Ciani, Clara, Clark, Cleva, Coccia, Cohadon, Colla, Collette, Colombini, Cominsky, Conte, Cook, Corbitt, Cordier, Cornish, Corpuz, Corsi, Costa, Coughlin, Coughlin, Coulon, Countryman, Couvares, Coward, Cowart, Coyne, Coyne, Craig, Creighton, Crowder, Cumming, Cunningham, Cuoco, Dahl, Canton, Damjanic, Danilishin, D’Antonio, Danzmann, Dattilo, Daveloza, Davier, Davies, Daw, Day, Dayanga, Debreczeni, Degallaix, Deléglise, Pozzo, Denker, Dent, Dereli, Dergachev, Rosa, DeRosa, DeSalvo, Dhurandhar, Díaz, Fiore, Lieto, Palma, Virgilio, Donath, Donovan, Dooley, Doravari, Dorosh, Dossa, Douglas, Downes, Drago, Drever, Driggers, Du, Dwyer, Eberle, Edo, Edwards, Effler, Eggenstein, Ehrens, Eichholz,
  Eikenberry, Endrőczi, Essick, Etzel, Evans, Evans, Factourovich, Fafone, Fairhurst, Fang, Farinon, Farr, Farr, Favata, Fehrmann, Fejer, Feldbaum, Feroz, Ferrante, Ferrini, Fidecaro, Finn, Fiori, Fisher, Flaminio, Fournier, Franco, Frasca, Frasconi, Frede, Frei, Freise, Frey, Fricke, Fritschel, Frolov, Fulda, Fyffe, Gair, Gammaitoni, Gaonkar, Garufi, Gehrels, Gemme, Genin, Gennai, Ghosh, Giaime, Giardina, Giazotto, Gill, Gleason, Goetz, Goetz, Gondan, González, Gordon, Gorodetsky, Gossan, Goßler, Gouaty, Gräf, Graff, Granata, Grant, Gras, Gray, Greenhalgh, Gretarsson, Groot, Grote, Grover, Grunewald, Guidi, Guido, Gushwa, Gustafson, Gustafson, Hammer, Hammond, Hanke, Hanks, Hanna, Hanson, Harms, Harry, Harry, Harstad, Hart, Hartman, Haster, Haughian, Heidmann, Heintze, Heitmann, Hello, Hemming, Hendry, Heng, Heptonstall, Heurs, Hewitson, Hild, Hoak, Hodge, Holt, Hooper, Hopkins, Hosken, Hough, Howell, Hu, Huerta, Hughey, Husa, Huttner, Huynh, Dinh, Ingram, Inta, Isogai, Ivanov, Iyer, Izumi, Jacobson,
  James, Jang, Jaranowski, Ji, Forteza, Johnson, Jones, Jones, Jonker, Ju, K, Kalmus, Kalogera, Kandhasamy, Kang, Kanner, Karlen, Kasprzack, Katsavounidis, Katzman, Kaufer, Kawabe, Kawazoe, Kéfélian, Keiser, Keitel, Kelley, Kells, Khalaidovski, Khalili, Khazanov, Kim, Kim, Kim, Kim, Kim, King, King, Kinzel, Kissel, Klimenko, Kline, Koehlenbeck, Kokeyama, Kondrashov, Koranda, Korth, Kowalska, Kozak, Kremin, Kringel, Krishnan, Królak, Kuehn, Kumar, Kumar, Kumar, Kuo, Kutynia, Kwee, Landry, Lantz, Larson, Lasky, Lawrie, Lazzarini, Lazzaro, Leaci, Leavey, Lebigot, Lee, Lee, Lee, Lee, Leonardi, Leong, Roux, Leroy, Letendre, Levin, Levine, Lewis, Li, Libbrecht, Libson, Lin, Littenberg, Litvine, Lockerbie, Lockett, Lodhia, Loew, Logue, Lombardi, Lorenzini, Loriette, Lormand, Losurdo, Lough, Lubinski, Lück, Luijten, Lundgren, Lynch, Ma, Macarthur, Macdonald, MacDonald, Machenschalk, MacInnis, Macleod, Sandoval, Mageswaran, Maglione, Mailand, Majorana, Maksimovic, Malvezzi, Man, Manca, Mandel, Mandic, Mangano,
  Mangini, Mantovani, Marchesoni, Marion, Márka, Márka, Markosyan, Maros, Marque, Martelli, Martin, Martin, Martinelli, Martynov, Marx, Mason, Masserot, Massinger, Matichard, Matone, Matzner, Mavalvala, Mazumder, Mazzolo, McCarthy, McClelland, McGuire, McIntyre, McIver, McLin, Meacher, Meadors, Mehmet, Meidam, Meinders, Melatos, Mendell, Mercer, Meshkov, Messenger, Meyers, Miao, Michel, Mikhailov, Milano, Milde, Miller, Minenkov, Mingarelli, Mishra, Mitra, Mitrofanov, Mitselmakher, Mittleman, Moe, Moesta, Mohan, Mohapatra, Moraru, Moreno, Morgado, Morriss, Mossavi, Mours, Lowry, Mueller, Mueller, Mukherjee, Mullavey, Munch, Murphy, Murray, Mytidis, Nagy, Kumar, Nardecchia, Naticchioni, Nayak, Necula, Nelemans, Neri, Neri, Newton, Nguyen, Nitz, Nocera, Nolting, Normandin, Nuttall, Ochsner, O’Dell, Oelker, Oh, Oh, Ohme, Oppermann, O’Reilly, O’Shaughnessy, Osthelder, Ottaway, Ottens, Overmier, Owen, Padilla, Pai, Palashov, Palomba, Pan, Pan, Pankow, Paoletti, Paoletti, Papa, Paris, Pasqualetti,
  Passaquieti, Passuello, Pedraza, Penn, Perreca, Phelps, Pichot, Pickenpack, Piergiovanni, Pierro, Pietka, Pinard, Pinto, Pitkin, Poeld, Poggiani, Poteomkin, Powell, Prasad, Premachandra, Prestegard, Price, Prijatelj, Privitera, Prodi, Prokhorov, Puncken, Punturo, Puppo, Qin, Quetschke, Quintero, Quiroga, James, Raab, Rabeling, Rácz, Radkins, Raffai, Raja, Rajalakshmi, Rakhmanov, Ramet, Ramirez, Rapagnani, Raymond, Re, Read, Reed, Regimbau, Reid, Reitze, Rhoades, Ricci, Riles, Robertson, Robinet, Rocchi, Rodruck, Rolland, Rollins, Romano, Romanov, Romie, Rosińska, Rowan, Rüdiger, Ruggi, Ryan, Salemi, Sammut, Sandberg, Sanders, Sannibale, Prieto, Saracco, Sassolas, Sathyaprakash, Saulson, Savage, Scheuer, Schilling, Schnabel, Schofield, Schreiber, Schuette, Schutz, Scott, Scott, Sellers, Sengupta, Sentenac, Sequino, Sergeev, Shaddock, Shah, Shahriar, Shaltev, Shapiro, Shawhan, Shoemaker, Sidery, Siellez, Siemens, Sigg, Simakov, Singer, Singer, Singh, Sintes, Slagmolen, Slutsky, Smith, Smith, Smith,
  Lefebvre, Son, Sorazu, Souradeep, Sperandio, Staley, Stebbins, Steinlechner, Steinlechner, Stephens, Steplewski, Stevenson, Stone, Stops, Strain, Straniero, Strigin, Sturani, Stuver, Summerscales, Susmithan, Sutton, Swinkels, Tacca, Talukder, Tanner, Tarabrin, Taylor, ter Braack, Thirugnanasambandam, Thomas, Thomas, Thorne, Thorne, Thrane, Tiwari, Tokmakov, Tomlinson, Toncelli, Tonelli, Torre, Torres, Torrie, Travasso, Traylor, Tse, Ugolini, Unnikrishnan, Urban, Urbanek, Vahlbruch, Vajente, Valdes, Vallisneri, vanden Brand, Broeck, vander Putten, vander Sluys, van Heijningen, van Veggel, Vass, Vasúth, Vaulin, Vecchio, Vedovato, Veitch, Veitch, Venkateswara, Verkindt, Verma, Vetrano, Viceré, Finley, Vinet, Vitale, Vo, Vocca, Vorvick, Vousden, Vyachanin, Wade, Wade, Wade, Walker, Wallace, Wang, Wang, Ward, Was, Weaver, Wei, Weinert, Weinstein, Weiss, Welborn, Wen, Wessels, West, Westphal, Wette, Whelan, White, Whiting, Wiesner, Wilkinson, Williams, Williams, Williams, Williams, Williamson, Willis, Willke,
  Wimmer, Winkler, Wipf, Wiseman, Wittel, Woan, Worden, Yablon, Yakushin, Yamamoto, Yancey, Yang, Yang, Yoshida, Yvert, Zadrożny, Zanolin, Zendri, Zhang, Zhang, Zhao, Zhu, Zucker, Zuraw, \& Zweizig}]{timedomainfstat2}
Aasi, J., Abbott, B.~P., Abbott, R., {et~al.} 2014, Classical and Quantum Gravity, 31, 165014, \dodoi{10.1088/0264-9381/31/16/165014}

\bibitem[{Aasi {et~al.}(2015)}]{AdvLIGOdetector}
Aasi, J., {et~al.} 2015, Class. Quant. Grav., 32, 074001, \dodoi{10.1088/0264-9381/32/7/074001}

\bibitem[{Abbott {et~al.}(2017{\natexlab{a}})}]{LIGOScientific:2017vwq}
Abbott, B.~P., {et~al.} 2017{\natexlab{a}}, Phys. Rev. Lett., 119, 161101, \dodoi{10.1103/PhysRevLett.119.161101}

\bibitem[{Abbott {et~al.}(2017{\natexlab{b}})}]{LIGOScientific:2017ync}
---. 2017{\natexlab{b}}, Astrophys. J. Lett., 848, L12, \dodoi{10.3847/2041-8213/aa91c9}

\bibitem[{Abbott {et~al.}(2018)}]{LIGOScientific:2018cki}
---. 2018, Phys. Rev. Lett., 121, 161101, \dodoi{10.1103/PhysRevLett.121.161101}

\bibitem[{Abbott {et~al.}(2020)}]{LIGOScientific:2020aai}
---. 2020, Astrophys. J. Lett., 892, L3, \dodoi{10.3847/2041-8213/ab75f5}

\bibitem[{Abbott {et~al.}(2021{\natexlab{a}})}]{LIGOScientific:2021djp}
Abbott, R., {et~al.} 2021{\natexlab{a}}.
\newblock \doarXiv{2111.03606}

\bibitem[{Abbott {et~al.}(2021{\natexlab{b}})}]{LIGOScientific:2021qlt}
---. 2021{\natexlab{b}}, Astrophys. J. Lett., 915, L5, \dodoi{10.3847/2041-8213/ac082e}

\bibitem[{Abbott {et~al.}(2021{\natexlab{c}})}]{LIGOScientific:2021sio}
---. 2021{\natexlab{c}}.
\newblock \doarXiv{2112.06861}

\bibitem[{Abbott {et~al.}(2022{\natexlab{a}})}]{KAGRA:2022osp}
---. 2022{\natexlab{a}}, Phys. Rev. D, 106, 042003, \dodoi{10.1103/PhysRevD.106.042003}

\bibitem[{Abbott {et~al.}(2022{\natexlab{b}})}]{KAGRA:2022dwb}
---. 2022{\natexlab{b}}, Phys. Rev. D, 106, 102008, \dodoi{10.1103/PhysRevD.106.102008}

\bibitem[{Abbott {et~al.}(2022{\natexlab{c}})}]{LIGOScientific:2021quq}
---. 2022{\natexlab{c}}, Astrophys. J., 932, 133, \dodoi{10.3847/1538-4357/ac6ad0}

\bibitem[{Abbott {et~al.}(2022{\natexlab{d}})Abbott, Abe, Acernese, Ackley, Adhikari, Adhikari, Adkins, Adya, Affeldt, Agarwal, Agathos, Agatsuma, Aggarwal, Aguiar, Aiello, Ain, Ajith, Akutsu, Albanesi, Alfaidi, Allocca, Altin, Amato, Anand, Anand, Ananyeva, Anderson, Anderson, Ando, Andrade, Andres, Andr\'es-Carcasona, Andri\ifmmode~\acute{c}\else \'{c}\fi{}, Angelova, Ansoldi, Antelis, Antier, Apostolatos, Appavuravther, Appert, Apple, Arai, Araya, Araya, Areeda, Ar\`ene, Aritomi, Arnaud, Arogeti, Aronson, Asada, Asali, Ashton, Aso, Assiduo, Assis~de Souza~Melo, Aston, Astone, Aubin, AultONeal, Austin, Babak, Badaracco, Bader, Badger, Bae, Bae, Baer, Bagnasco, Bai, Baird, Bajpai, Baka, Ball, Ballardin, Ballmer, Balsamo, Baltus, Banagiri, Banerjee, Bankar, Barayoga, Barbieri, Barish, Barker, Barneo, Barone, Barr, Barsotti, Barsuglia, Barta, Bartlett, Barton, Bartos, Basak, Bassiri, Basti, Bawaj, Bayley, Bazzan, Becher, B\'ecsy, Bedakihale, Beirnaert, Bejger, Belahcene, Benedetto, Beniwal, Benjamin, Bennett,
  Bentley, BenYaala, Bera, Berbel, Bergamin, Berger, Bernuzzi, Bersanetti, Bertolini, Betzwieser, Beveridge, Bhandare, Bhandari, Bhardwaj, Bhatt, Bhattacharjee, Bhaumik, Bianchi, Bilenko, Billingsley, Bini, Birney, Birnholtz, Biscans, Bischi, Biscoveanu, Bisht, Biswas, Bitossi, Bizouard, Blackburn, Blair, Blair, Blair, Bobba, Bode, Bo\"er, Bogaert, Boldrini, Bolingbroke, Bonavena, Bondu, Bonilla, Bonnand, Booker, Boom, Bork, Boschi, Bose, Bose, Bossilkov, Boudart, Bouffanais, Bozzi, Bradaschia, Brady, Bramley, Branch, Branchesi, Brau, Breschi, Briant, Briggs, Brillet, Brinkmann, Brockill, Brooks, Brooks, Brown, Brunett, Bruno, Bruntz, Bryant, Bucci, Bulik, Bulten, Buonanno, Burtnyk, Buscicchio, Buskulic, Buy, Byer, Cabourn~Davies, Cabras, Cabrita, Cadonati, Caesar, Cagnoli, Cahillane, Calder\'on~Bustillo, Callaghan, Callister, Calloni, Cameron, Camp, Canepa, Canevarolo, Cannavacciuolo, Cannon, Cao, Cao, Capocasa, Capote, Carapella, Carbognani, Carlassara, Carlin, Carney, Carpinelli, Carrillo, Carullo, Carver,
  Casanueva~Diaz, Casentini, Castaldi, Caudill, Cavagli\`a, Cavalier, Cavalieri, Cella, Cerd\'a-Dur\'an, Cesarini, Chaibi, Chalathadka~Subrahmanya, Champion, Chan, Chan, Chan, Chan, Chan, Chandra, Chang, Chanial, Chao, Chapman-Bird, Charlton, Chase, Chassande-Mottin, Chatterjee, Chatterjee, Chatterjee, Chaturvedi, Chaty, Chen, Chen, Chen, Chen, Chen, Chen, Chen, Chen, Chen, Cheng, Cheong, Cheung, Chia, Chiadini, Chiang, Chiarini, Chierici, Chincarini, Chiofalo, Chiummo, Choudhary, Choudhary, Christensen, Chu, Chu, Chua, Chung, Ciani, Ciecielag, Cie\ifmmode~\acute{s}\else \'{s}\fi{}lar, Cifaldi, Ciobanu, Ciolfi, Cipriano, Clara, Clark, Clearwater, Clesse, Cleva, Coccia, Codazzo, Cohadon, Cohen, Colleoni, Collette, Colombo, Colpi, Compton, Constancio, Conti, Cooper, Corban, Corbitt, Cordero-Carri\'on, Corezzi, Corley, Cornish, Corre, Corsi, Cortese, Costa, Cotesta, Cottingham, Coughlin, Coulon, Countryman, Cousins, Couvares, Coward, Cowart, Coyne, Coyne, Creighton, Creighton, Criswell, Croquette, Crowder,
  Cudell, Cullen, Cumming, Cummings, Cunningham, Cuoco, Cury\l{}o, Dabadie, Dal~Canton, Dall'Osso, D\'alya, Dana, D'Angelo, Danilishin, D'Antonio, Danzmann, Darsow-Fromm, Dasgupta, Datrier, Datta, Datta, Dattilo, Dave, Davier, Davis, Davis, Daw, Dean, DeBra, Deenadayalan, Degallaix, De~Laurentis, Del\'eglise, Del~Favero, De~Lillo, De~Lillo, Dell'Aquila, Del~Pozzo, DeMarchi, De~Matteis, D'Emilio, Demos, Dent, Depasse, De~Pietri, De~Rosa, De~Rossi, DeSalvo, De~Simone, Dhurandhar, D\'{\i}az, Di~Cesare, Didio, Dietrich, Di~Fiore, Di~Fronzo, Di~Giorgio, Di~Giovanni, Di~Giovanni, Di~Girolamo, Di~Lieto, Di~Michele, Ding, Di~Pace, Di~Palma, Di~Renzo, Divakarla, Dmitriev, Doctor, Donahue, D'Onofrio, Donovan, Dooley, Doravari, Dorosh, Drago, Driggers, Drori, Ducoin, Dupej, Dupletsa, Durante, D'Urso, Duverne, Dwyer, Eassa, Easter, Ebersold, Eckhardt, Eddolls, Edelman, Edo, Edy, Effler, Eguchi, Eichholz, Eikenberry, Eisenmann, Eisenstein, Ejlli, Engelby, Enomoto, Errico, Essick, Estell\'es, Estevez, Etienne, Etzel,
  Evans, Evans, Evstafyeva, Ewing, Fabrizi, Faedi, Fafone, Fair, Fairhurst, Fan, Farah, Farinon, Farr, Farr, Fauchon-Jones, Favaro, Favata, Fays, Fazio, Feicht, Fejer, Fenyvesi, Ferguson, Fernandez-Galiana, Ferrante, Ferreira, Fidecaro, Figura, Fiori, Fiori, Fishbach, Fisher, Fittipaldi, Fiumara, Flaminio, Floden, Fong, Font, Fornal, Forsyth, Franke, Frasca, Frasconi, Freed, Frei, Freise, Freitas, Frey, Fritschel, Frolov, Fronz\'e, Fujii, Fujikawa, Fujimoto, Fulda, Fyffe, Gabbard, Gadre, Gair, Gais, Galaudage, Gamba, Ganapathy, Ganguly, Gao, Gaonkar, Garaventa, Garc\'{\i}a N\'u\~nez, Garc\'{\i}a-Quir\'os, Garufi, Gateley, Gayathri, Ge, Gemme, Gennai, George, Gerberding, Gergely, Gewecke, Ghonge, Ghosh, Ghosh, Ghosh, Ghosh, Ghosh, Giacomazzo, Giacoppo, Giaime, Giardina, Gibson, Gier, Giesler, Giri, Gissi, Gkaitatzis, Glanzer, Gleckl, Godwin, Goetz, Goetz, Gohlke, Golomb, Goncharov, Gonz\'alez, Gosselin, Gouaty, Gould, Goyal, Grace, Grado, Graham, Granata, Granata, Grant, Gras, Grassia, Gray, Gray, Greco,
  Green, Green, Gretarsson, Gretarsson, Griffith, Griffiths, Griggs, Grignani, Grimaldi, Grimes, Grimm, Grote, Grunewald, Gruning, Gruson, Guerra, Guidi, Guimaraes, Guix\'e, Gulati, Gunny, Guo, Guo, Gupta, Gupta, Gupta, Gupta, Gupta, Gustafson, Guzman, Ha, Hadiputrawan, Haegel, Haino, Halim, Hall, Hamilton, Hammond, Han, Haney, Hanks, Hanna, Hannam, Hannuksela, Hansen, Hansen, Hanson, Harder, Haris, Harms, Harry, Harry, Hartwig, Hasegawa, Haskell, Haster, Hathaway, Hattori, Haughian, Hayakawa, Hayama, Hayes, Healy, Heidmann, Heidt, Heintze, Heinze, Heinzel, Heitmann, Hellman, Hello, Helmling-Cornell, Hemming, Hendry, Heng, Hennes, Hennig, Hennig, Henshaw, Hernandez, Hernandez~Vivanco, Heurs, Hewitt, Higginbotham, Hild, Hill, Himemoto, Hines, Hirata, Hirose, Ho, Hochheim, Hofman, Hohmann, Holcomb, Holland, Hollows, Holmes, Holt, Holz, Hong, Hough, Hourihane, Howell, Hoy, Hoyland, Hreibi, Hsieh, Hsieh, Hsiung, Hsu, Huang, Huang, Huang, Huang, Huang, Huang, H\"ubner, Huddart, Hughey, Hui, Hui, Husa, Huttner,
  Huxford, Huynh-Dinh, Ide, Idzkowski, Iess, Inayoshi, Inoue, Iosif, Isi, Isleif, Ito, Itoh, Iyer, JaberianHamedan, Jacqmin, Jacquet, Jadhav, Jadhav, Jain, James, Jan, Jani, Janquart, Janssens, Janthalur, Jaranowski, Jariwala, Jaume, Jenkins, Jenner, Jeon, Jia, Jiang, Jin, Johns, Johnston, Jones, Jones, Jones, Jones, Joshi, Ju, Jue, Jung, Jung, Junker, Juste, Kaihotsu, Kajita, Kakizaki, Kalaghatgi, Kalogera, Kamai, Kamiizumi, Kanda, Kandhasamy, Kang, Kanner, Kao, Kapadia, Kapasi, Karathanasis, Karki, Kashyap, Kasprzack, Kastaun, Kato, Katsanevas, Katsavounidis, Katzman, Kaur, Kawabe, Kawaguchi, K\'ef\'elian, Keitel, Key, Khadka, Khalili, Khan, Khanam, Khazanov, Khetan, Khursheed, Kijbunchoo, Kim, Kim, Kim, Kim, Kim, Kim, Kim, Kimball, Kimura, Kinley-Hanlon, Kirchhoff, Kissel, Klimenko, Klinger, Knee, Knowles, Knust, Knyazev, Kobayashi, Koch, Koekoek, Kohri, Kokeyama, Koley, Kolitsidou, Kolstein, Komori, Kondrashov, Kong, Kontos, Koper, Korobko, Kovalam, Koyama, Kozak, Kozakai, Kringel, Krishnendu, Kr\'olak,
  Kuehn, Kuei, Kuijer, Kulkarni, Kumar, Kumar, Kumar, Kumar, Kume, Kuns, Kuromiya, Kuroyanagi, Kwak, Lacaille, Lagabbe, Laghi, Lalande, Lalleman, Lam, Lamberts, Landry, Lane, Lang, Lange, Lantz, La~Rosa, Lartaux-Vollard, Lasky, Laxen, Lazzarini, Lazzaro, Leaci, Leavey, LeBohec, Lecoeuche, Lee, Lee, Lee, Lee, Lee, Legred, Lehmann, Lema\^{\i}tre, Lenti, Leonardi, Leonova, Leroy, Letendre, Levesque, Levin, Leviton, Leyde, Li, Li, Li, Li, Li, Li, Li, Lin, Lin, Lin, Lin, Lin, Lin, Linde, Linker, Linley, Littenberg, Liu, Liu, Liu, Liu, Llamas, Lo, Lo, London, Longo, Lopez, Lopez~Portilla, Lorenzini, Loriette, Lormand, Losurdo, Lott, Lough, Lousto, Lovelace, Lucaccioni, L\"uck, Lumaca, Lundgren, Luo, Lynam, Ma'arif, Macas, Machtinger, MacInnis, Macleod, MacMillan, Macquet, Maga\~na Hernandez, Magazz\`u, Magee, Maggiore, Magnozzi, Mahesh, Majorana, Maksimovic, Maliakal, Malik, Man, Mandic, Mangano, Mansell, Manske, Mantovani, Mapelli, Marchesoni, Mar\'{\i}n~Pina, Marion, Mark, M\'arka, M\'arka, Markakis, Markosyan,
  Markowitz, Maros, Marquina, Marsat, Martelli, Martin, Martin, Martinez, Martinez, Martinez, Martinovic, Martynov, Marx, Masalehdan, Mason, Massera, Masserot, Masso-Reid, Mastrogiovanni, Matas, Mateu-Lucena, Matichard, Matiushechkina, Mavalvala, McCann, McCarthy, McClelland, McClincy, McCormick, McCuller, McGhee, McGuire, McIsaac, McIver, McRae, McWilliams, Meacher, Mehmet, Mehta, Meijer, Melatos, Melchor, Mendell, Menendez-Vazquez, Menoni, Mercer, Mereni, Merfeld, Merilh, Merritt, Merzougui, Meshkov, Messenger, Messick, Meyers, Meylahn, Mhaske, Miani, Miao, Michaloliakos, Michel, Michimura, Middleton, Mihaylov, Milano, Miller, Miller, Miller, Millhouse, Mills, Milotti, Minenkov, Mio, Mir, Miravet-Ten\'es, Mishkin, Mishra, Mishra, Mistry, Mitra, Mitrofanov, Mitselmakher, Mittleman, Miyakawa, Miyo, Miyoki, Mo, Modafferi, Moguel, Mogushi, Mohapatra, Mohite, Molina, Molina-Ruiz, Mondin, Montani, Moore, Moragues, Moraru, Morawski, More, Moreno, Moreno, Mori, Morisaki, Morisue, Moriwaki, Mours, Mow-Lowry, Mozzon,
  Muciaccia, Mukherjee, Mukherjee, Mukherjee, Mukherjee, Mukherjee, Mukund, Mullavey, Munch, Mu\~niz, Murray, Musenich, Muusse, Nadji, Nagano, Nagar, Nakamura, Nakano, Nakano, Nakayama, Napolano, Nardecchia, Narikawa, Narola, Naticchioni, Nayak, Nayak, Neil, Neilson, Nelson, Nelson, Nery, Neubauer, Neunzert, Ng, Ng, Nguyen, Nguyen, Nguyen, Quynh, Ni, Ni, Nichols, Nishimoto, Nishizawa, Nissanke, Nitoglia, Nocera, Norman, North, Nozaki, Nurbek, Nuttall, Obayashi, Oberling, O'Brien, O'Dell, Oelker, Ogaki, Oganesyan, Oh, Oh, Oh, Ohashi, Ohashi, Ohkawa, Ohme, Ohta, Okada, Okutani, Olivetto, Oohara, Oram, O'Reilly, Ormiston, Ormsby, O'Shaughnessy, O'Shea, Oshino, Ossokine, Osthelder, Otabe, Ottaway, Overmier, Pace, Pagano, Pagano, Page, Pagliaroli, Pai, Pai, Pal, Palamos, Palashov, Palomba, Pan, Pan, Panda, Pang, Pankow, Pannarale, Pant, Panther, Paoletti, Paoli, Paolone, Pappas, Parisi, Park, Park, Parker, Pascucci, Pasqualetti, Passaquieti, Passuello, Patel, Pathak, Patricelli, Patron, Paul, Payne, Pedraza,
  Pedurand, Pegoraro, Pele, Pe\~na Arellano, Penano, Penn, Perego, Pereira, Pereira, Perez, P\'erigois, Perkins, Perreca, Perri\`es, Pesios, Petermann, Petterson, Pfeiffer, Pham, Pham, Phukon, Phurailatpam, Piccinni, Pichot, Piendibene, Piergiovanni, Pierini, Pierro, Pillant, Pillas, Pilo, Pinard, Pineda-Bosque, Pinto, Pinto, Piotrzkowski, Piotrzkowski, Pirello, Pisarski, Pitkin, Placidi, Placidi, Planas, Plastino, Pluchar, Poggiani, Polini, Pong, Ponrathnam, Porter, Poulton, Poverman, Powell, Pracchia, Pradier, Prajapati, Prasai, Prasanna, Pratten, Principe, Prodi, Prokhorov, Prosposito, Prudenzi, Puecher, Punturo, Puosi, Puppo, P\"urrer, Qi, Quartey, Quetschke, Quinonez, Quitzow-James, Raab, Raaijmakers, Radkins, Radulesco, Raffai, Rail, Raja, Rajan, Ramirez, Ramirez, Ramos-Buades, Rana, Rapagnani, Ray, Raymond, Raza, Razzano, Read, Rees, Regimbau, Rei, Reid, Reid, Reitze, Relton, Renzini, Rettegno, Revenu, Reza, Rezac, Ricci, Richards, Richardson, Richardson, Riemenschneider, Riles, Rinaldi, Rink,
  Robertson, Robie, Robinet, Rocchi, Rodriguez, Rolland, Rollins, Romanelli, Romano, Romel, Romero, Romero-Shaw, Romie, Ronchini, Rosa, Rose, Rosi\ifmmode~\acute{n}\else \'{n}\fi{}ska, Ross, Rowan, Rowlinson, Roy, Roy, Roy, Rozza, Ruggi, Ruiz-Rocha, Ryan, Sachdev, Sadecki, Sadiq, Saha, Saito, Sakai, Sakellariadou, Sakon, Salafia, Salces-Carcoba, Salconi, Saleem, Salemi, Samajdar, Sanchez, Sanchez, Sanchez, Sanchis-Gual, Sanders, Sanuy, Saravanan, Sarin, Sassolas, Satari, Sauter, Savage, Savant, Sawada, Sawant, Sayah, Schaetzl, Scheel, Scheuer, Schiworski, Schmidt, Schmidt, Schnabel, Schneewind, Schofield, Sch\"onbeck, Schulte, Schutz, Schwartz, Scott, Scott, Seglar-Arroyo, Sekiguchi, Sellers, Sengupta, Sentenac, Seo, Sequino, Sergeev, Setyawati, Shaffer, Shahriar, Shaikh, Shams, Shao, Sharma, Sharma, Shawhan, Shcheblanov, Sheela, Shikano, Shikauchi, Shimizu, Shimode, Shinkai, Shishido, Shoda, Shoemaker, Shoemaker, ShyamSundar, Sieniawska, Sigg, Silenzi, Singer, Singh, Singh, Singh, Singha, Sintes, Sipala,
  Skliris, Slagmolen, Slaven-Blair, Smetana, Smith, Smith, Smith, Soldateschi, Somala, Somiya, Song, Soni, Soni, Sordini, Sorrentino, Sorrentino, Soulard, Souradeep, Sowell, Spagnuolo, Spencer, Spera, Spinicelli, Srivastava, Srivastava, Staats, Stachie, Stachurski, Steer, Steinlechner, Steinlechner, Stergioulas, Stops, Stover, Strain, Strang, Stratta, Strong, Strunk, Sturani, Stuver, Suchenek, Sudhagar, Sudhir, Sugimoto, Suh, Sullivan, Summerscales, Sun, Sunil, Sur, Suresh, Sutton, Suzuki, Suzuki, Suzuki, Swinkels, Szczepa\ifmmode~\acute{n}\else \'{n}\fi{}czyk, Szewczyk, Tacca, Tagoshi, Tait, Takahashi, Takahashi, Takano, Takeda, Takeda, Talbot, Talbot, Tanaka, Tanaka, Tanaka, Tanasijczuk, Tanioka, Tanner, Tao, Tao, Tapia, Tapia San~Mart\'{\i}n, Taranto, Taruya, Tasson, Tenorio, Terhune, Terkowski, Thirugnanasambandam, Thomas, Thomas, Thompson, Thompson, Thondapu, Thorne, Thrane, Tiwari, Tiwari, Tiwari, Toivonen, Tolley, Tomaru, Tomura, Tonelli, Tornasi, Torres-Forn\'e, Torrie, Tosta~e Melo, T\"oyr\"a,
  Trapananti, Travasso, Traylor, Trevor, Tringali, Tripathee, Troiano, Trovato, Trozzo, Trudeau, Tsai, Tsang, Tsang, Tsao, Tse, Tso, Tsuchida, Tsukada, Tsuna, Tsutsui, Turbang, Turconi, Tuyenbayev, Ubhi, Uchikata, Uchiyama, Udall, Ueda, Uehara, Ueno, Ueshima, Unnikrishnan, Urban, Ushiba, Utina, Vajente, Vajpeyi, Valdes, Valentini, Valsan, van Bakel, van Beuzekom, van Dael, van~den Brand, Van Den~Broeck, Vander-Hyde, van Haevermaet, van Heijningen, van Putten, van Remortel, Vardaro, Vargas, Varma, Vas\'uth, Vecchio, Vedovato, Veitch, Veitch, Venneberg, Venugopalan, Verkindt, Verma, Verma, Vermeulen, Veske, Vetrano, Vicer\'e, Vidyant, Viets, Vijaykumar, Villa-Ortega, Vinet, Virtuoso, Vitale, Vocca, von Reis, von Wrangel, Vorvick, Vyatchanin, Wade, Wade, Wagner, Walet, Walker, Wallace, Wallace, Wang, Wang, Wang, Ward, Warner, Was, Washimi, Washington, Watchi, Weaver, Weaving, Webster, Weinert, Weinstein, Weiss, Weller, Weller, Wellmann, Wen, We\ss{}els, Wette, Whelan, White, Whiting, Whittle, Wilken, Williams,
  Williams, Williamson, Willis, Willke, Wilson, Wipf, Wlodarczyk, Woan, Woehler, Wofford, Wong, Wong, Wright, Wu, Wu, Wu, Wysocki, Xiao, Yamada, Yamamoto, Yamamoto, Yamamoto, Yamashita, Yamazaki, Yang, Yang, Yang, Yang, Yang, Yang, Yap, Yeeles, Yeh, Yelikar, Ying, Yokoyama, Yokozawa, Yoo, Yoshioka, Yu, Yu, Yuzurihara, Zadro\ifmmode~\dot{z}\else \.{z}\fi{}ny, Zanolin, Zeidler, Zelenova, Zendri, Zevin, Zhan, Zhang, Zhang, Zhang, Zhang, Zhang, Zhang, Zhao, Zhao, Zhao, Zhao, Zhou, Zhou, Zhu, Zhu, Zucker, \& Zweizig}]{upperlimitdata}
Abbott, R., Abe, H., Acernese, F., {et~al.} 2022{\natexlab{d}}, All-sky search for continuous gravitational waves from isolated neutron stars using Advanced LIGO and Advanced Virgo O3 data.
\newblock \url{https://dcc.ligo.org/LIGO-P2100367/public}

\bibitem[{Abbott {et~al.}(2022{\natexlab{e}})Abbott, Abe, Acernese, Ackley, Adhikari, Adhikari, Adkins, Adya, Affeldt, Agarwal, Agathos, Agatsuma, Aggarwal, Aguiar, Aiello, Ain, Ajith, Akutsu, Albanesi, Alfaidi, Allocca, Altin, Amato, Anand, Anand, Ananyeva, Anderson, Anderson, Ando, Andrade, Andres, Andr\'es-Carcasona, Andri\ifmmode~\acute{c}\else \'{c}\fi{}, Angelova, Ansoldi, Antelis, Antier, Apostolatos, Appavuravther, Appert, Apple, Arai, Araya, Araya, Areeda, Ar\`ene, Aritomi, Arnaud, Arogeti, Aronson, Asada, Asali, Ashton, Aso, Assiduo, Assis~de Souza~Melo, Aston, Astone, Aubin, AultONeal, Austin, Babak, Badaracco, Bader, Badger, Bae, Bae, Baer, Bagnasco, Bai, Baird, Bajpai, Baka, Ball, Ballardin, Ballmer, Balsamo, Baltus, Banagiri, Banerjee, Bankar, Barayoga, Barbieri, Barish, Barker, Barneo, Barone, Barr, Barsotti, Barsuglia, Barta, Bartlett, Barton, Bartos, Basak, Bassiri, Basti, Bawaj, Bayley, Bazzan, Becher, B\'ecsy, Bedakihale, Beirnaert, Bejger, Belahcene, Benedetto, Beniwal, Benjamin, Bennett,
  Bentley, BenYaala, Bera, Berbel, Bergamin, Berger, Bernuzzi, Bersanetti, Bertolini, Betzwieser, Beveridge, Bhandare, Bhandari, Bhardwaj, Bhatt, Bhattacharjee, Bhaumik, Bianchi, Bilenko, Billingsley, Bini, Birney, Birnholtz, Biscans, Bischi, Biscoveanu, Bisht, Biswas, Bitossi, Bizouard, Blackburn, Blair, Blair, Blair, Bobba, Bode, Bo\"er, Bogaert, Boldrini, Bolingbroke, Bonavena, Bondu, Bonilla, Bonnand, Booker, Boom, Bork, Boschi, Bose, Bose, Bossilkov, Boudart, Bouffanais, Bozzi, Bradaschia, Brady, Bramley, Branch, Branchesi, Brau, Breschi, Briant, Briggs, Brillet, Brinkmann, Brockill, Brooks, Brooks, Brown, Brunett, Bruno, Bruntz, Bryant, Bucci, Bulik, Bulten, Buonanno, Burtnyk, Buscicchio, Buskulic, Buy, Byer, Cabourn~Davies, Cabras, Cabrita, Cadonati, Caesar, Cagnoli, Cahillane, Calder\'on~Bustillo, Callaghan, Callister, Calloni, Cameron, Camp, Canepa, Canevarolo, Cannavacciuolo, Cannon, Cao, Cao, Capocasa, Capote, Carapella, Carbognani, Carlassara, Carlin, Carney, Carpinelli, Carrillo, Carullo, Carver,
  Casanueva~Diaz, Casentini, Castaldi, Caudill, Cavagli\`a, Cavalier, Cavalieri, Cella, Cerd\'a-Dur\'an, Cesarini, Chaibi, Chalathadka~Subrahmanya, Champion, Chan, Chan, Chan, Chan, Chan, Chandra, Chang, Chanial, Chao, Chapman-Bird, Charlton, Chase, Chassande-Mottin, Chatterjee, Chatterjee, Chatterjee, Chaturvedi, Chaty, Chen, Chen, Chen, Chen, Chen, Chen, Chen, Chen, Chen, Cheng, Cheong, Cheung, Chia, Chiadini, Chiang, Chiarini, Chierici, Chincarini, Chiofalo, Chiummo, Choudhary, Choudhary, Christensen, Chu, Chu, Chua, Chung, Ciani, Ciecielag, Cie\ifmmode~\acute{s}\else \'{s}\fi{}lar, Cifaldi, Ciobanu, Ciolfi, Cipriano, Clara, Clark, Clearwater, Clesse, Cleva, Coccia, Codazzo, Cohadon, Cohen, Colleoni, Collette, Colombo, Colpi, Compton, Constancio, Conti, Cooper, Corban, Corbitt, Cordero-Carri\'on, Corezzi, Corley, Cornish, Corre, Corsi, Cortese, Costa, Cotesta, Cottingham, Coughlin, Coulon, Countryman, Cousins, Couvares, Coward, Cowart, Coyne, Coyne, Creighton, Creighton, Criswell, Croquette, Crowder,
  Cudell, Cullen, Cumming, Cummings, Cunningham, Cuoco, Cury\l{}o, Dabadie, Dal~Canton, Dall'Osso, D\'alya, Dana, D'Angelo, Danilishin, D'Antonio, Danzmann, Darsow-Fromm, Dasgupta, Datrier, Datta, Datta, Dattilo, Dave, Davier, Davis, Davis, Daw, Dean, DeBra, Deenadayalan, Degallaix, De~Laurentis, Del\'eglise, Del~Favero, De~Lillo, De~Lillo, Dell'Aquila, Del~Pozzo, DeMarchi, De~Matteis, D'Emilio, Demos, Dent, Depasse, De~Pietri, De~Rosa, De~Rossi, DeSalvo, De~Simone, Dhurandhar, D\'{\i}az, Di~Cesare, Didio, Dietrich, Di~Fiore, Di~Fronzo, Di~Giorgio, Di~Giovanni, Di~Giovanni, Di~Girolamo, Di~Lieto, Di~Michele, Ding, Di~Pace, Di~Palma, Di~Renzo, Divakarla, Dmitriev, Doctor, Donahue, D'Onofrio, Donovan, Dooley, Doravari, Dorosh, Drago, Driggers, Drori, Ducoin, Dupej, Dupletsa, Durante, D'Urso, Duverne, Dwyer, Eassa, Easter, Ebersold, Eckhardt, Eddolls, Edelman, Edo, Edy, Effler, Eguchi, Eichholz, Eikenberry, Eisenmann, Eisenstein, Ejlli, Engelby, Enomoto, Errico, Essick, Estell\'es, Estevez, Etienne, Etzel,
  Evans, Evans, Evstafyeva, Ewing, Fabrizi, Faedi, Fafone, Fair, Fairhurst, Fan, Farah, Farinon, Farr, Farr, Fauchon-Jones, Favaro, Favata, Fays, Fazio, Feicht, Fejer, Fenyvesi, Ferguson, Fernandez-Galiana, Ferrante, Ferreira, Fidecaro, Figura, Fiori, Fiori, Fishbach, Fisher, Fittipaldi, Fiumara, Flaminio, Floden, Fong, Font, Fornal, Forsyth, Franke, Frasca, Frasconi, Freed, Frei, Freise, Freitas, Frey, Fritschel, Frolov, Fronz\'e, Fujii, Fujikawa, Fujimoto, Fulda, Fyffe, Gabbard, Gadre, Gair, Gais, Galaudage, Gamba, Ganapathy, Ganguly, Gao, Gaonkar, Garaventa, Garc\'{\i}a N\'u\~nez, Garc\'{\i}a-Quir\'os, Garufi, Gateley, Gayathri, Ge, Gemme, Gennai, George, Gerberding, Gergely, Gewecke, Ghonge, Ghosh, Ghosh, Ghosh, Ghosh, Ghosh, Giacomazzo, Giacoppo, Giaime, Giardina, Gibson, Gier, Giesler, Giri, Gissi, Gkaitatzis, Glanzer, Gleckl, Godwin, Goetz, Goetz, Gohlke, Golomb, Goncharov, Gonz\'alez, Gosselin, Gouaty, Gould, Goyal, Grace, Grado, Graham, Granata, Granata, Grant, Gras, Grassia, Gray, Gray, Greco,
  Green, Green, Gretarsson, Gretarsson, Griffith, Griffiths, Griggs, Grignani, Grimaldi, Grimes, Grimm, Grote, Grunewald, Gruning, Gruson, Guerra, Guidi, Guimaraes, Guix\'e, Gulati, Gunny, Guo, Guo, Gupta, Gupta, Gupta, Gupta, Gupta, Gustafson, Guzman, Ha, Hadiputrawan, Haegel, Haino, Halim, Hall, Hamilton, Hammond, Han, Haney, Hanks, Hanna, Hannam, Hannuksela, Hansen, Hansen, Hanson, Harder, Haris, Harms, Harry, Harry, Hartwig, Hasegawa, Haskell, Haster, Hathaway, Hattori, Haughian, Hayakawa, Hayama, Hayes, Healy, Heidmann, Heidt, Heintze, Heinze, Heinzel, Heitmann, Hellman, Hello, Helmling-Cornell, Hemming, Hendry, Heng, Hennes, Hennig, Hennig, Henshaw, Hernandez, Hernandez~Vivanco, Heurs, Hewitt, Higginbotham, Hild, Hill, Himemoto, Hines, Hirata, Hirose, Ho, Hochheim, Hofman, Hohmann, Holcomb, Holland, Hollows, Holmes, Holt, Holz, Hong, Hough, Hourihane, Howell, Hoy, Hoyland, Hreibi, Hsieh, Hsieh, Hsiung, Hsu, Huang, Huang, Huang, Huang, Huang, Huang, H\"ubner, Huddart, Hughey, Hui, Hui, Husa, Huttner,
  Huxford, Huynh-Dinh, Ide, Idzkowski, Iess, Inayoshi, Inoue, Iosif, Isi, Isleif, Ito, Itoh, Iyer, JaberianHamedan, Jacqmin, Jacquet, Jadhav, Jadhav, Jain, James, Jan, Jani, Janquart, Janssens, Janthalur, Jaranowski, Jariwala, Jaume, Jenkins, Jenner, Jeon, Jia, Jiang, Jin, Johns, Johnston, Jones, Jones, Jones, Jones, Joshi, Ju, Jue, Jung, Jung, Junker, Juste, Kaihotsu, Kajita, Kakizaki, Kalaghatgi, Kalogera, Kamai, Kamiizumi, Kanda, Kandhasamy, Kang, Kanner, Kao, Kapadia, Kapasi, Karathanasis, Karki, Kashyap, Kasprzack, Kastaun, Kato, Katsanevas, Katsavounidis, Katzman, Kaur, Kawabe, Kawaguchi, K\'ef\'elian, Keitel, Key, Khadka, Khalili, Khan, Khanam, Khazanov, Khetan, Khursheed, Kijbunchoo, Kim, Kim, Kim, Kim, Kim, Kim, Kim, Kimball, Kimura, Kinley-Hanlon, Kirchhoff, Kissel, Klimenko, Klinger, Knee, Knowles, Knust, Knyazev, Kobayashi, Koch, Koekoek, Kohri, Kokeyama, Koley, Kolitsidou, Kolstein, Komori, Kondrashov, Kong, Kontos, Koper, Korobko, Kovalam, Koyama, Kozak, Kozakai, Kringel, Krishnendu, Kr\'olak,
  Kuehn, Kuei, Kuijer, Kulkarni, Kumar, Kumar, Kumar, Kumar, Kume, Kuns, Kuromiya, Kuroyanagi, Kwak, Lacaille, Lagabbe, Laghi, Lalande, Lalleman, Lam, Lamberts, Landry, Lane, Lang, Lange, Lantz, La~Rosa, Lartaux-Vollard, Lasky, Laxen, Lazzarini, Lazzaro, Leaci, Leavey, LeBohec, Lecoeuche, Lee, Lee, Lee, Lee, Lee, Legred, Lehmann, Lema\^{\i}tre, Lenti, Leonardi, Leonova, Leroy, Letendre, Levesque, Levin, Leviton, Leyde, Li, Li, Li, Li, Li, Li, Li, Lin, Lin, Lin, Lin, Lin, Lin, Linde, Linker, Linley, Littenberg, Liu, Liu, Liu, Liu, Llamas, Lo, Lo, London, Longo, Lopez, Lopez~Portilla, Lorenzini, Loriette, Lormand, Losurdo, Lott, Lough, Lousto, Lovelace, Lucaccioni, L\"uck, Lumaca, Lundgren, Luo, Lynam, Ma'arif, Macas, Machtinger, MacInnis, Macleod, MacMillan, Macquet, Maga\~na Hernandez, Magazz\`u, Magee, Maggiore, Magnozzi, Mahesh, Majorana, Maksimovic, Maliakal, Malik, Man, Mandic, Mangano, Mansell, Manske, Mantovani, Mapelli, Marchesoni, Mar\'{\i}n~Pina, Marion, Mark, M\'arka, M\'arka, Markakis, Markosyan,
  Markowitz, Maros, Marquina, Marsat, Martelli, Martin, Martin, Martinez, Martinez, Martinez, Martinovic, Martynov, Marx, Masalehdan, Mason, Massera, Masserot, Masso-Reid, Mastrogiovanni, Matas, Mateu-Lucena, Matichard, Matiushechkina, Mavalvala, McCann, McCarthy, McClelland, McClincy, McCormick, McCuller, McGhee, McGuire, McIsaac, McIver, McRae, McWilliams, Meacher, Mehmet, Mehta, Meijer, Melatos, Melchor, Mendell, Menendez-Vazquez, Menoni, Mercer, Mereni, Merfeld, Merilh, Merritt, Merzougui, Meshkov, Messenger, Messick, Meyers, Meylahn, Mhaske, Miani, Miao, Michaloliakos, Michel, Michimura, Middleton, Mihaylov, Milano, Miller, Miller, Miller, Millhouse, Mills, Milotti, Minenkov, Mio, Mir, Miravet-Ten\'es, Mishkin, Mishra, Mishra, Mistry, Mitra, Mitrofanov, Mitselmakher, Mittleman, Miyakawa, Miyo, Miyoki, Mo, Modafferi, Moguel, Mogushi, Mohapatra, Mohite, Molina, Molina-Ruiz, Mondin, Montani, Moore, Moragues, Moraru, Morawski, More, Moreno, Moreno, Mori, Morisaki, Morisue, Moriwaki, Mours, Mow-Lowry, Mozzon,
  Muciaccia, Mukherjee, Mukherjee, Mukherjee, Mukherjee, Mukherjee, Mukund, Mullavey, Munch, Mu\~niz, Murray, Musenich, Muusse, Nadji, Nagano, Nagar, Nakamura, Nakano, Nakano, Nakayama, Napolano, Nardecchia, Narikawa, Narola, Naticchioni, Nayak, Nayak, Neil, Neilson, Nelson, Nelson, Nery, Neubauer, Neunzert, Ng, Ng, Nguyen, Nguyen, Nguyen, Quynh, Ni, Ni, Nichols, Nishimoto, Nishizawa, Nissanke, Nitoglia, Nocera, Norman, North, Nozaki, Nurbek, Nuttall, Obayashi, Oberling, O'Brien, O'Dell, Oelker, Ogaki, Oganesyan, Oh, Oh, Oh, Ohashi, Ohashi, Ohkawa, Ohme, Ohta, Okada, Okutani, Olivetto, Oohara, Oram, O'Reilly, Ormiston, Ormsby, O'Shaughnessy, O'Shea, Oshino, Ossokine, Osthelder, Otabe, Ottaway, Overmier, Pace, Pagano, Pagano, Page, Pagliaroli, Pai, Pai, Pal, Palamos, Palashov, Palomba, Pan, Pan, Panda, Pang, Pankow, Pannarale, Pant, Panther, Paoletti, Paoli, Paolone, Pappas, Parisi, Park, Park, Parker, Pascucci, Pasqualetti, Passaquieti, Passuello, Patel, Pathak, Patricelli, Patron, Paul, Payne, Pedraza,
  Pedurand, Pegoraro, Pele, Pe\~na Arellano, Penano, Penn, Perego, Pereira, Pereira, Perez, P\'erigois, Perkins, Perreca, Perri\`es, Pesios, Petermann, Petterson, Pfeiffer, Pham, Pham, Phukon, Phurailatpam, Piccinni, Pichot, Piendibene, Piergiovanni, Pierini, Pierro, Pillant, Pillas, Pilo, Pinard, Pineda-Bosque, Pinto, Pinto, Piotrzkowski, Piotrzkowski, Pirello, Pisarski, Pitkin, Placidi, Placidi, Planas, Plastino, Pluchar, Poggiani, Polini, Pong, Ponrathnam, Porter, Poulton, Poverman, Powell, Pracchia, Pradier, Prajapati, Prasai, Prasanna, Pratten, Principe, Prodi, Prokhorov, Prosposito, Prudenzi, Puecher, Punturo, Puosi, Puppo, P\"urrer, Qi, Quartey, Quetschke, Quinonez, Quitzow-James, Raab, Raaijmakers, Radkins, Radulesco, Raffai, Rail, Raja, Rajan, Ramirez, Ramirez, Ramos-Buades, Rana, Rapagnani, Ray, Raymond, Raza, Razzano, Read, Rees, Regimbau, Rei, Reid, Reid, Reitze, Relton, Renzini, Rettegno, Revenu, Reza, Rezac, Ricci, Richards, Richardson, Richardson, Riemenschneider, Riles, Rinaldi, Rink,
  Robertson, Robie, Robinet, Rocchi, Rodriguez, Rolland, Rollins, Romanelli, Romano, Romel, Romero, Romero-Shaw, Romie, Ronchini, Rosa, Rose, Rosi\ifmmode~\acute{n}\else \'{n}\fi{}ska, Ross, Rowan, Rowlinson, Roy, Roy, Roy, Rozza, Ruggi, Ruiz-Rocha, Ryan, Sachdev, Sadecki, Sadiq, Saha, Saito, Sakai, Sakellariadou, Sakon, Salafia, Salces-Carcoba, Salconi, Saleem, Salemi, Samajdar, Sanchez, Sanchez, Sanchez, Sanchis-Gual, Sanders, Sanuy, Saravanan, Sarin, Sassolas, Satari, Sauter, Savage, Savant, Sawada, Sawant, Sayah, Schaetzl, Scheel, Scheuer, Schiworski, Schmidt, Schmidt, Schnabel, Schneewind, Schofield, Sch\"onbeck, Schulte, Schutz, Schwartz, Scott, Scott, Seglar-Arroyo, Sekiguchi, Sellers, Sengupta, Sentenac, Seo, Sequino, Sergeev, Setyawati, Shaffer, Shahriar, Shaikh, Shams, Shao, Sharma, Sharma, Shawhan, Shcheblanov, Sheela, Shikano, Shikauchi, Shimizu, Shimode, Shinkai, Shishido, Shoda, Shoemaker, Shoemaker, ShyamSundar, Sieniawska, Sigg, Silenzi, Singer, Singh, Singh, Singh, Singha, Sintes, Sipala,
  Skliris, Slagmolen, Slaven-Blair, Smetana, Smith, Smith, Smith, Soldateschi, Somala, Somiya, Song, Soni, Soni, Sordini, Sorrentino, Sorrentino, Soulard, Souradeep, Sowell, Spagnuolo, Spencer, Spera, Spinicelli, Srivastava, Srivastava, Staats, Stachie, Stachurski, Steer, Steinlechner, Steinlechner, Stergioulas, Stops, Stover, Strain, Strang, Stratta, Strong, Strunk, Sturani, Stuver, Suchenek, Sudhagar, Sudhir, Sugimoto, Suh, Sullivan, Summerscales, Sun, Sunil, Sur, Suresh, Sutton, Suzuki, Suzuki, Suzuki, Swinkels, Szczepa\ifmmode~\acute{n}\else \'{n}\fi{}czyk, Szewczyk, Tacca, Tagoshi, Tait, Takahashi, Takahashi, Takano, Takeda, Takeda, Talbot, Talbot, Tanaka, Tanaka, Tanaka, Tanasijczuk, Tanioka, Tanner, Tao, Tao, Tapia, Tapia San~Mart\'{\i}n, Taranto, Taruya, Tasson, Tenorio, Terhune, Terkowski, Thirugnanasambandam, Thomas, Thomas, Thompson, Thompson, Thondapu, Thorne, Thrane, Tiwari, Tiwari, Tiwari, Toivonen, Tolley, Tomaru, Tomura, Tonelli, Tornasi, Torres-Forn\'e, Torrie, Tosta~e Melo, T\"oyr\"a,
  Trapananti, Travasso, Traylor, Trevor, Tringali, Tripathee, Troiano, Trovato, Trozzo, Trudeau, Tsai, Tsang, Tsang, Tsao, Tse, Tso, Tsuchida, Tsukada, Tsuna, Tsutsui, Turbang, Turconi, Tuyenbayev, Ubhi, Uchikata, Uchiyama, Udall, Ueda, Uehara, Ueno, Ueshima, Unnikrishnan, Urban, Ushiba, Utina, Vajente, Vajpeyi, Valdes, Valentini, Valsan, van Bakel, van Beuzekom, van Dael, van~den Brand, Van Den~Broeck, Vander-Hyde, van Haevermaet, van Heijningen, van Putten, van Remortel, Vardaro, Vargas, Varma, Vas\'uth, Vecchio, Vedovato, Veitch, Veitch, Venneberg, Venugopalan, Verkindt, Verma, Verma, Vermeulen, Veske, Vetrano, Vicer\'e, Vidyant, Viets, Vijaykumar, Villa-Ortega, Vinet, Virtuoso, Vitale, Vocca, von Reis, von Wrangel, Vorvick, Vyatchanin, Wade, Wade, Wagner, Walet, Walker, Wallace, Wallace, Wang, Wang, Wang, Ward, Warner, Was, Washimi, Washington, Watchi, Weaver, Weaving, Webster, Weinert, Weinstein, Weiss, Weller, Weller, Wellmann, Wen, We\ss{}els, Wette, Whelan, White, Whiting, Whittle, Wilken, Williams,
  Williams, Williamson, Willis, Willke, Wilson, Wipf, Wlodarczyk, Woan, Woehler, Wofford, Wong, Wong, Wright, Wu, Wu, Wu, Wysocki, Xiao, Yamada, Yamamoto, Yamamoto, Yamamoto, Yamashita, Yamazaki, Yang, Yang, Yang, Yang, Yang, Yang, Yap, Yeeles, Yeh, Yelikar, Ying, Yokoyama, Yokozawa, Yoo, Yoshioka, Yu, Yu, Yuzurihara, Zadro\ifmmode~\dot{z}\else \.{z}\fi{}ny, Zanolin, Zeidler, Zelenova, Zendri, Zevin, Zhan, Zhang, Zhang, Zhang, Zhang, Zhang, Zhang, Zhao, Zhao, Zhao, Zhao, Zhou, Zhou, Zhu, Zhu, Zucker, \& Zweizig}]{allskyo3}
---. 2022{\natexlab{e}}, Phys. Rev. D, 106, 102008, \dodoi{10.1103/PhysRevD.106.102008}

\bibitem[{Abbott {et~al.}(2023{\natexlab{a}})}]{KAGRA:2021duu}
Abbott, R., {et~al.} 2023{\natexlab{a}}, Phys. Rev. X, 13, 011048, \dodoi{10.1103/PhysRevX.13.011048}

\bibitem[{Abbott {et~al.}(2023{\natexlab{b}})}]{LIGOScientific:2021aug}
---. 2023{\natexlab{b}}, Astrophys. J., 949, 76, \dodoi{10.3847/1538-4357/ac74bb}

\bibitem[{Acernese {et~al.}(2015)}]{AdvVIRGOdetector}
Acernese, F., {et~al.} 2015, Class. Quant. Grav., 32, 024001, \dodoi{10.1088/0264-9381/32/2/024001}

\bibitem[{Akutsu {et~al.}(2021)}]{KAGRAdetector}
Akutsu, T., {et~al.} 2021, PTEP, 2021, 05A101, \dodoi{10.1093/ptep/ptaa125}

\bibitem[{Astone {et~al.}(2014)Astone, Colla, D'Antonio, Frasca, \& Palomba}]{frequencyhough}
Astone, P., Colla, A., D'Antonio, S., Frasca, S., \& Palomba, C. 2014, Phys. Rev. D, 90, 042002, \dodoi{10.1103/PhysRevD.90.042002}

\bibitem[{{Astropy Collaboration} {et~al.}(2013){Astropy Collaboration}, {Robitaille}, {Tollerud}, {Greenfield}, {Droettboom}, {Bray}, {Aldcroft}, {Davis}, {Ginsburg}, {Price-Whelan}, {Kerzendorf}, {Conley}, {Crighton}, {Barbary}, {Muna}, {Ferguson}, {Grollier}, {Parikh}, {Nair}, {Unther}, {Deil}, {Woillez}, {Conseil}, {Kramer}, {Turner}, {Singer}, {Fox}, {Weaver}, {Zabalza}, {Edwards}, {Azalee Bostroem}, {Burke}, {Casey}, {Crawford}, {Dencheva}, {Ely}, {Jenness}, {Labrie}, {Lim}, {Pierfederici}, {Pontzen}, {Ptak}, {Refsdal}, {Servillat}, \& {Streicher}}]{2013A&A...558A..33A}
{Astropy Collaboration}, {Robitaille}, T.~P., {Tollerud}, E.~J., {et~al.} 2013, \aap, 558, A33, \dodoi{10.1051/0004-6361/201322068}

\bibitem[{{Astropy Collaboration} {et~al.}(2018){Astropy Collaboration}, {Price-Whelan}, {Sip{\H{o}}cz}, {G{\"u}nther}, {Lim}, {Crawford}, {Conseil}, {Shupe}, {Craig}, {Dencheva}, {Ginsburg}, {VanderPlas}, {Bradley}, {P{\'e}rez-Su{\'a}rez}, {de Val-Borro}, {Aldcroft}, {Cruz}, {Robitaille}, {Tollerud}, {Ardelean}, {Babej}, {Bach}, {Bachetti}, {Bakanov}, {Bamford}, {Barentsen}, {Barmby}, {Baumbach}, {Berry}, {Biscani}, {Boquien}, {Bostroem}, {Bouma}, {Brammer}, {Bray}, {Breytenbach}, {Buddelmeijer}, {Burke}, {Calderone}, {Cano Rodr{\'\i}guez}, {Cara}, {Cardoso}, {Cheedella}, {Copin}, {Corrales}, {Crichton}, {D'Avella}, {Deil}, {Depagne}, {Dietrich}, {Donath}, {Droettboom}, {Earl}, {Erben}, {Fabbro}, {Ferreira}, {Finethy}, {Fox}, {Garrison}, {Gibbons}, {Goldstein}, {Gommers}, {Greco}, {Greenfield}, {Groener}, {Grollier}, {Hagen}, {Hirst}, {Homeier}, {Horton}, {Hosseinzadeh}, {Hu}, {Hunkeler}, {Ivezi{\'c}}, {Jain}, {Jenness}, {Kanarek}, {Kendrew}, {Kern}, {Kerzendorf}, {Khvalko}, {King}, {Kirkby}, {Kulkarni},
  {Kumar}, {Lee}, {Lenz}, {Littlefair}, {Ma}, {Macleod}, {Mastropietro}, {McCully}, {Montagnac}, {Morris}, {Mueller}, {Mumford}, {Muna}, {Murphy}, {Nelson}, {Nguyen}, {Ninan}, {N{\"o}the}, {Ogaz}, {Oh}, {Parejko}, {Parley}, {Pascual}, {Patil}, {Patil}, {Plunkett}, {Prochaska}, {Rastogi}, {Reddy Janga}, {Sabater}, {Sakurikar}, {Seifert}, {Sherbert}, {Sherwood-Taylor}, {Shih}, {Sick}, {Silbiger}, {Singanamalla}, {Singer}, {Sladen}, {Sooley}, {Sornarajah}, {Streicher}, {Teuben}, {Thomas}, {Tremblay}, {Turner}, {Terr{\'o}n}, {van Kerkwijk}, {de la Vega}, {Watkins}, {Weaver}, {Whitmore}, {Woillez}, {Zabalza}, \& {Astropy Contributors}}]{2018AJ....156..123A}
{Astropy Collaboration}, {Price-Whelan}, A.~M., {Sip{\H{o}}cz}, B.~M., {et~al.} 2018, \aj, 156, 123, \dodoi{10.3847/1538-3881/aabc4f}

\bibitem[{Bayley {et~al.}(2019)Bayley, Messenger, \& Woan}]{soap}
Bayley, J., Messenger, C., \& Woan, G. 2019, Phys. Rev. D, 100, 023006, \dodoi{10.1103/PhysRevD.100.023006}

\bibitem[{Bonazzola \& Gourgoulhon(1996{\natexlab{a}})}]{Bonazzola:1995rb}
Bonazzola, S., \& Gourgoulhon, E. 1996{\natexlab{a}}, Astron. Astrophys., 312, 675.
\newblock \doarXiv{astro-ph/9602107}

\bibitem[{Bonazzola \& Gourgoulhon(1996{\natexlab{b}})}]{bonazzola1996gravitational}
---. 1996{\natexlab{b}}, arXiv preprint astro-ph/9602107

\bibitem[{Brown \& Bildsten(1998)}]{brown1998ocean}
Brown, E.~F., \& Bildsten, L. 1998, The Astrophysical Journal, 496, 915

\bibitem[{Chandrasekhar \& Fermi(1953)}]{chandrasekhar1953problems}
Chandrasekhar, S., \& Fermi, E. 1953, Astrophysical Journal, 116

\bibitem[{Dergachev \& Papa(2023)}]{dergachev2023frequency}
Dergachev, V., \& Papa, M.~A. 2023, Physical Review X, 13, 021020

\bibitem[{Dergachev \& Papa(2024)}]{dergachev2024early}
---. 2024, Physical Review D, 109, 022007

\bibitem[{Ferraro(1954)}]{ferraro1954equilibrium}
Ferraro, V. 1954, Astrophysical Journal, vol. 119, p. 407, 119, 407

\bibitem[{Gal'tsov \& Tsvetkov(1984)}]{gal1984gravitational}
Gal'tsov, D., \& Tsvetkov, V. 1984, Physics Letters A, 103, 193

\bibitem[{Glampedakis \& Gualtieri(2018)}]{glampedakis2018}
Glampedakis, K., \& Gualtieri, L. 2018, The Physics and Astrophysics of Neutron Stars, 673

\bibitem[{Glampedakis {et~al.}(2012)Glampedakis, Jones, \& Samuelsson}]{glampedakis2012}
Glampedakis, K., Jones, D., \& Samuelsson, L. 2012, Physical review letters, 109, 081103

\bibitem[{Glendenning(2012)}]{glendenning2012compact}
Glendenning, N.~K. 2012, Compact stars: Nuclear physics, particle physics and general relativity (Springer Science \& Business Media)

\bibitem[{Haensel \& Zdunik(1990)}]{haensel1990non}
Haensel, P., \& Zdunik, J. 1990, Astronomy and Astrophysics (ISSN 0004-6361), vol. 227, no. 2, Jan. 1990, p. 431-436., 227, 431

\bibitem[{Haskell {et~al.}(2007)Haskell, Andersson, Jones, \& Samuelsson}]{haskell2007}
Haskell, B., Andersson, N., Jones, D., \& Samuelsson, L. 2007, Physical Review Letters, 99, 231101

\bibitem[{Haskell {et~al.}(2006)Haskell, Jones, \& Andersson}]{haskell2006mountains}
Haskell, B., Jones, D., \& Andersson, N. 2006, Monthly Notices of the Royal Astronomical Society, 373, 1423

\bibitem[{Haskell {et~al.}(2008)Haskell, Samuelsson, Glampedakis, \& Andersson}]{haskell2008modelling}
Haskell, B., Samuelsson, L., Glampedakis, K., \& Andersson, N. 2008, Monthly Notices of the Royal Astronomical Society, 385, 531

\bibitem[{Hunter(2007)}]{Hunter:2007}
Hunter, J.~D. 2007, Computing in Science \& Engineering, 9, 90, \dodoi{10.1109/MCSE.2007.55}

\bibitem[{Jaranowski {et~al.}(1998{\natexlab{a}})Jaranowski, Kr\'olak, \& Schutz}]{jaranowski1998}
Jaranowski, P., Kr\'olak, A., \& Schutz, B.~F. 1998{\natexlab{a}}, Phys. Rev. D, 58, 063001, \dodoi{10.1103/PhysRevD.58.063001}

\bibitem[{Jaranowski {et~al.}(1998{\natexlab{b}})Jaranowski, Kr\'olak, \& Schutz}]{timedomainfstat1}
---. 1998{\natexlab{b}}, Phys. Rev. D, 58, 063001, \dodoi{10.1103/PhysRevD.58.063001}

\bibitem[{{Johnson-McDaniel} \& {Owen}(2013)}]{nathan2013}
{Johnson-McDaniel}, N.~K., \& {Owen}, B.~J. 2013, \prd, 88, 044004, \dodoi{10.1103/PhysRevD.88.044004}

\bibitem[{Kluyver {et~al.}(2016)Kluyver, Ragan-Kelley, P{\'e}rez, Granger, Bussonnier, Frederic, Kelley, Hamrick, Grout, Corlay, Ivanov, Avila, Abdalla, Willing, \& development team}]{jupyter}
Kluyver, T., Ragan-Kelley, B., P{\'e}rez, F., {et~al.} 2016, in Positioning and Power in Academic Publishing: Players, Agents and Agendas, ed. F.~Loizides \& B.~Scmidt (Netherlands: IOS Press), 87--90.
\newblock \url{https://eprints.soton.ac.uk/403913/}

\bibitem[{Knispel \& Allen(2008)}]{knispel2008}
Knispel, B., \& Allen, B. 2008, Phys. Rev. D, 78, 044031, \dodoi{10.1103/PhysRevD.78.044031}

\bibitem[{Konno {et~al.}(2000)Konno, Obata, \& Kojima}]{konno2000flattening}
Konno, K., Obata, T., \& Kojima, Y. 2000, arXiv preprint astro-ph/0001397

\bibitem[{Krishnan {et~al.}(2004)Krishnan, Sintes, Papa, Schutz, Frasca, \& Palomba}]{skyhough}
Krishnan, B., Sintes, A.~M., Papa, M.~A., {et~al.} 2004, Phys. Rev. D, 70, 082001, \dodoi{10.1103/PhysRevD.70.082001}

\bibitem[{Lasky(2015)}]{lasky2015gravitational}
Lasky, P.~D. 2015, Publications of the Astronomical Society of Australia, 32, e034

\bibitem[{Manchester {et~al.}(2005)Manchester, Hobbs, Teoh, \& Hobbs}]{manchester2005}
Manchester, R.~N., Hobbs, G.~B., Teoh, A., \& Hobbs, M. 2005, The Astronomical Journal, 129, 1993

\bibitem[{Mastrano {et~al.}(2011)Mastrano, Melatos, Reisenegger, \& Akg{\"u}n}]{mastrano2011gravitational}
Mastrano, A., Melatos, A., Reisenegger, A., \& Akg{\"u}n, T. 2011, Monthly Notices of the Royal Astronomical Society, 417, 2288

\bibitem[{Melatos \& Payne(2005)}]{melatos2005gravitational}
Melatos, A., \& Payne, D. 2005, The Astrophysical Journal, 623, 1044

\bibitem[{Morales \& Horowitz(2022)}]{morales2022}
Morales, J., \& Horowitz, C. 2022, Monthly Notices of the Royal Astronomical Society, 517, 5610

\bibitem[{Olausen \& Kaspi(2014)}]{olausen2014mcgill}
Olausen, S., \& Kaspi, V. 2014, The Astrophysical Journal Supplement Series, 212, 6

\bibitem[{Owen(2005)}]{Owen:2005fn}
Owen, B.~J. 2005, Phys. Rev. Lett., 95, 211101, \dodoi{10.1103/PhysRevLett.95.211101}

\bibitem[{{Paczynski}(1990)}]{Paczynski}
{Paczynski}, B. 1990, \apj, 348, 485, \dodoi{10.1086/168257}

\bibitem[{Pagliaro {et~al.}(2023)Pagliaro, Papa, Ming, Lian, Tsuna, Maraston, \& Thomas}]{pagliaro2023continuous}
Pagliaro, G., Papa, M.~A., Ming, J., {et~al.} 2023, The Astrophysical Journal, 952, 123, \dodoi{10.3847/1538-4357/acd76f}

\bibitem[{Piccinni(2022)}]{Piccinni:2022vsd}
Piccinni, O.~J. 2022, Galaxies, 10, 72, \dodoi{10.3390/galaxies10030072}

\bibitem[{Punturo {et~al.}(2010)Punturo, Abernathy, Acernese, Allen, Andersson, Arun, Barone, Barr, Barsuglia, Beker, {et~al.}}]{punturo2010}
Punturo, M., Abernathy, M., Acernese, F., {et~al.} 2010, Classical and Quantum Gravity, 27, 194002

\bibitem[{{Reed} {et~al.}(2021){Reed}, {Deibel}, \& {Horowitz}}]{Reed}
{Reed}, B.~T., {Deibel}, A., \& {Horowitz}, C.~J. 2021, \apj, 921, 89, \dodoi{10.3847/1538-4357/ac1c04}

\bibitem[{Reitze {et~al.}(2019)}]{Reitze:2019iox}
Reitze, D., {et~al.} 2019, Bull. Am. Astron. Soc., 51, 035.
\newblock \doarXiv{1907.04833}

\bibitem[{Riles(2023)}]{Riles:2022wwz}
Riles, K. 2023, Living Rev. Rel., 26, 3, \dodoi{10.1007/s41114-023-00044-3}

\bibitem[{Steltner {et~al.}(2023)Steltner, Papa, Eggenstein, Prix, Bensch, Allen, \& Machenschalk}]{steltner2023deep}
Steltner, B., Papa, M., Eggenstein, H.-B., {et~al.} 2023, The Astrophysical Journal, 952, 55

\bibitem[{Tenorio(2023)}]{tenorio2023blindsearch}
Tenorio, R. 2023, Blind-search constraints on the sub-kiloparsec population of continuous gravitational-wave sources.
\newblock \doarXiv{2310.12097}

\bibitem[{Treves {et~al.}(2000)Treves, Turolla, Zane, \& Colpi}]{treves2000isolated}
Treves, A., Turolla, R., Zane, S., \& Colpi, M. 2000, Publications of the Astronomical Society of the Pacific, 112, 297

\bibitem[{Ushomirsky {et~al.}(2000)Ushomirsky, Cutler, \& Bildsten}]{ushomirsky2000deformations}
Ushomirsky, G., Cutler, C., \& Bildsten, L. 2000, Monthly Notices of the Royal Astronomical Society, 319, 902

\bibitem[{van~der Walt {et~al.}(2011)van~der Walt, Colbert, \& Varoquaux}]{vanderWalt:2011bqk}
van~der Walt, S., Colbert, S.~C., \& Varoquaux, G. 2011, Comput. Sci. Eng., 13, 22, \dodoi{10.1109/MCSE.2011.37}

\bibitem[{Virtanen {et~al.}(2020)}]{Virtanen:2019joe}
Virtanen, P., {et~al.} 2020, Nature Meth., \dodoi{10.1038/s41592-019-0686-2}

\bibitem[{Wade {et~al.}(2012)Wade, Siemens, Kaplan, Knispel, \& Allen}]{wade}
Wade, L., Siemens, X., Kaplan, D.~L., Knispel, B., \& Allen, B. 2012, Phys. Rev. D, 86, 124011, \dodoi{10.1103/PhysRevD.86.124011}

\bibitem[{Weber {et~al.}(2007)Weber, Negreiros, Rosenfield, \& Stejner}]{weber2007pulsars}
Weber, F., Negreiros, R., Rosenfield, P., \& Stejner, M. 2007, Progress in Particle and Nuclear Physics, 59, 94

\bibitem[{{W}es {M}c{K}inney(2010)}]{mckinney-proc-scipy-2010}
{W}es {M}c{K}inney. 2010, in {P}roceedings of the 9th {P}ython in {S}cience {C}onference, ed. {S}t\'efan van~der {W}alt \& {J}arrod {M}illman, 56 -- 61, \dodoi{10.25080/Majora-92bf1922-00a}

\bibitem[{Xu(2003)}]{xu2003solid}
Xu, R. 2003, The Astrophysical Journal, 596, L59

\end{thebibliography}
\bibliographystyle{aasjournal}



\end{document}